\documentclass[letterpaper,USenglish]{lipics-v2021}

\hideLIPIcs  %
\nolinenumbers
\pdfoutput=1

\usepackage{graphicx}
\usepackage{xcolor}
\usepackage{url}
\usepackage{algorithm}
\usepackage{algpseudocode}
\usepackage{caption}
\usepackage{subcaption}
\usepackage{xspace}
\usepackage{listings}
\definecolor{KWColor}{rgb}{0.37,0.08,0.25}
\definecolor{CommentColor}{rgb}{0.12,0.38,0.18}
\definecolor{StringColor}{rgb}{0.06,0.10,0.98}
\definecolor{darkred}{rgb}{0.75,0,0}
\definecolor{lightgrey}{rgb}{0.8,0.8,0.8}

\lstdefinestyle{Eclipse}{
  xleftmargin=0pt,
  basicstyle=\ttfamily\footnotesize,
  commentstyle=\color{CommentColor}\ttfamily\small,
  stringstyle=\color{StringColor},
  keywordstyle=\color{KWColor}\bfseries,
  escapeinside={/*@}{@*/}
}

\lstdefinelanguage{JavaScript}[]{Java}{
   morekeywords={debugger,delete,function,in,typeof,var,with},
   morestring=[b]'
}

\newcommand{\code}[1]{\lstinline[basicstyle=\ttfamily\small]{#1}}

\usepackage{array}
\newcolumntype{C}[1]{>{\centering\arraybackslash}m{#1}}

\newcommand{\codify}[1]{\texttt{\small #1}}
\newcommand{\tightpara}[1]{\subparagraph*{#1}}
  {\list{}{\leftmargin=0.15in\rightmargin=0.15in}\item[]}%
  {\endlist}

\usepackage{hyperref}

\usepackage[noabbrev]{cleveref}
\crefname{algocf}{Algorithm}{Algorithms}
\Crefname{algocf}{Algorithm}{Algorithms}

\usepackage{etoolbox}
\newtoggle{extended}
\toggletrue{extended}

\usepackage[nocompress,space]{cite}
\title{Automatic Root Cause Quantification for Missing Edges in JavaScript Call Graphs (Extended Version)}
\titlerunning{Automatic Root Cause Quantification for Missing Edges in JavaScript Call Graphs}

\author{Madhurima Chakraborty}{University of California, Riverside}{mchak009@ucr.edu}{}{}
\author{Renzo Olivares}{University of California, Riverside}{roliv006@ucr.edu}{}{}
\author{Manu Sridharan}{University of California, Riverside}{roliv006@ucr.edu}{}{}
\author{Behnaz Hassanshahi}{Oracle Labs Australia}{behnaz.hassanshahi@oracle.com}{}{}
\authorrunning{M. Chakraborty et al.}
\Copyright{Madhurima Chakraborty, Renzo Olivares, Manu Sridharan, and Behnaz Hassanshahi}



\funding{This research was supported
  in part by a gift from Oracle Labs and by the National Science Foundation under grant CCF-2007024.  This research was partially sponsored by the OUSD(R\&E)/RT\&L and was accomplished under
Cooperative Agreement Number W911NF-20-2-0267. The views and conclusions contained in this document are
those of the authors and should not be interpreted as representing the official policies, either expressed or implied, of
the ARL and OUSD(R\&E)/RT\&L or the U.S. Government. The U.S. Government is authorized to reproduce and
distribute reprints for Government purposes notwithstanding any copyright notation herein.}%

\EventEditors{Karim Ali and Jan Vitek}
\EventNoEds{2}
\EventLongTitle{36th European Conference on Object-Oriented Programming (ECOOP 2022)}
\EventShortTitle{ECOOP 2022}
\EventAcronym{ECOOP}
\EventYear{2022}
\EventDate{June 6--10, 2022}
\EventLocation{Berlin, Germany}
\EventLogo{}
\SeriesVolume{222}
\ArticleNo{3}

\begin{document}

\maketitle

\begin{abstract}

  Building sound and precise static call graphs for real-world JavaScript applications poses an enormous challenge, due to many hard-to-analyze language features.  Further, the relative importance of these features may vary depending on the call graph algorithm being used and the class of applications being analyzed.  In this paper, we present a technique to \emph{automatically} quantify the relative importance of different root causes of call graph unsoundness for a set of target applications.  The technique works by identifying the dynamic function data flows relevant to each call edge missed by the static analysis, correctly handling cases with multiple root causes and inter-dependent calls.  We apply our approach to perform a detailed study of the recall of a state-of-the-art call graph construction technique on a set of framework-based web applications.  The study yielded a number of useful insights.  We found that while dynamic property accesses were the most common root cause of missed edges across the benchmarks, other root causes varied in importance depending on the benchmark, potentially useful information for an analysis designer.  Further, with our approach, we could quickly identify and fix a recall issue in the call graph builder we studied, and also quickly assess whether a recent analysis technique for Node.js-based applications would be helpful for browser-based code.  All of our code and data is publicly available, and many components of our technique can be re-used to facilitate future studies.

\end{abstract}

\section{Introduction}\label{sec:intro}

Effective call graph construction is critically important for JavaScript static analysis, as JavaScript analysis tools often need to reason about behaviors that span function boundaries (e.g., security vulnerabilities~\cite{hassanshahi22gelato,GuarnieriPTDTB11} or correctness of library updates~\cite{moller20detecting}).  Unfortunately, call graph construction for real-world JavaScript programs poses significant challenges, particularly for client-side code in web applications.  Modern web applications are increasingly built using sophisticated frameworks like React~\cite{ReactJS} and AngularJS~\cite{AngularJS}.\footnote{A recent Stack Overflow developer survey shows popularity of these frameworks is growing, with total usage surpassing older libraries like jQuery~\cite{stackoverflowsurvey}.}  Sophisticated recent JavaScript static analysis frameworks~\cite{JensenMT09,safe,Kashyap2014,SchaferSDT13} often focus on sound and precise handling of complex JavaScript constructs.  While these systems have advanced significantly, they cannot yet scale to handle modern web frameworks.  There are also a growing number of unsound but pragmatic call graph analyses designed primarily to give useful results for real-world code bases~\cite{Feldthaus2013,moller20detecting,codeqlcg,nielsen21modular}.  While these techniques have been shown effective in certain domains, their unsoundness can lead to missing many edges when analyzing framework-based applications~\cite{hassanshahi22gelato}, i.e., the analyses can have low \emph{recall}.  For bug-finding and security analyses, these missing edges are of key concern as they can lead to false negatives like missed vulnerabilities.

To guide development of better call graph builders, it would be highly useful to know which language constructs are contributing most to reducing recall for a set of benchmarks of interest.  JavaScript has many different constructs that are typically ignored or only partially handled by pragmatic static analyses, due to their dynamic nature~\cite{Richards2010}.  Further, there are complex tradeoffs involved in adding support for these constructs, as a more complete handling may lead to scalability and precision problems.  Analysis designers aiming to improve results for a set of benchmarks would be helped by quantitative guidance on the relative importance of different unhandled language features.

\begin{figure}
  \centering
  \includegraphics[width=0.9\linewidth,keepaspectratio]{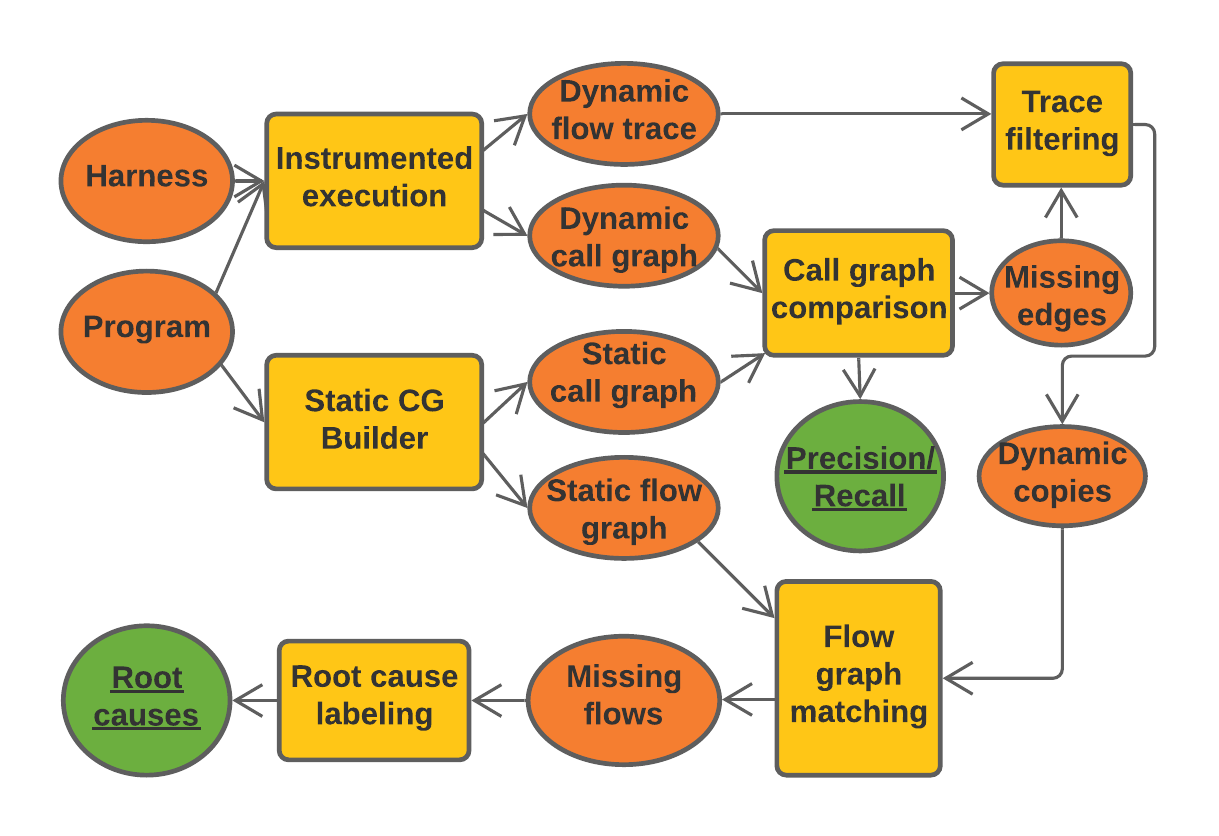}
  \caption{Overview of our methodology.}
  \label{fig:methodology-overview}
\end{figure}

This paper presents a novel technique for \emph{automatic root cause quantification} for missing edges in JavaScript call graphs.  \cref{fig:methodology-overview} gives an overview of our technique.  Given a program, a static call graph builder enhanced to also export static flow graphs (see \Cref{sec:cg-construction}), and a harness for exercising the program, our technique automatically finds \emph{missing flows}, data flows of function values that occur at runtime but are not modeled by the static analysis.  Our technique associates a set of missing flows with each missed call graph edge, thereby indicating which data flows must be handled by the static analysis to discover the missed edge.  The technique correctly accounts for \emph{inter-dependent calls}, where a call graph edge is missing due to the absence of other call graph edges.

We further observe that given a missing flow, one can often automatically determine a \emph{root cause label} for the flow, indicating which unhandled language construct(s) were responsible for the flow being missed.  Such labeling can be performed at different levels of granularity, depending on what level of detail is desired by the analysis designer.  Given logic to map missing flows to root cause labels, our technique automatically quantifies the prevalence of each root cause for the desired benchmarks.  

We have implemented our techniques, and we used them to study the recall of two variants of the approximate call graphs (ACG) algorithm of Feldthaus et al.~\cite{Feldthaus2013}, as implemented in the WALA framework~\cite{wala}, on a suite of modern web applications.  We found the root cause quantification to provide useful insights, in particular:
\begin{itemize}
  \item To our surprise, a large initial cause of low recall was the lack of models in WALA for a variety of built-in library functions.  By adding models, we were able to increase recall by up to 5 percentage points.
  \item After fixing the native models, dynamic property accesses were the largest root cause of low recall, at 70\%.  The second-largest root cause varied significantly across the benchmarks.
  \item We applied a finer-grained root cause labeling for dynamic property accesses, and found that their property names are computed in a wide variety of ways, with no single dominant pattern.  We studied the potential of a recently-described recall-improving technique for dynamic property accesses in Node.js programs~\cite{nielsen21modular}, and found that it would at best have a small impact for our web-based benchmarks.
\end{itemize}

Our dynamic call graph and flow trace analyses were challenging to implement due to JavaScript's hard-to-analyze language features.  JavaScript includes many difficult-to-analyze features, including (but not limited to) reflective call mechanisms, ``native'' library methods, getter/setter methods, and dynamic code evaluation.  Pragmatic static analyses often ignore most of these features, as they do not aim for sound results.  However, since we aimed to study which calls were missed by such analyses and \emph{why} those calls were missed, our dynamic analyses had to faithfully capture the behavior of these features, and thereby incurred significant additional complexity (see \cref{sec:flow-graph-matching}).

All of our code and data is publicly available in an artifact~\cite{acgArtifact}.  Our infrastructure is reusable and could be applied to study other static analyses, other benchmarks, and other platforms (e.g., Node.js).  Together, our infrastructure, methodology, and results can help guide the design of future analyses targeting real-world JavaScript code.

\subparagraph*{Contributions} This paper makes the following contributions:
\begin{itemize}
  \item We present a novel approach to quantifying the importance of language features causing low recall in JavaScript call graphs.  The approach properly handles missing call graph edges with multiple root causes, and also inter-dependent calls, where an edge is missing due to the absence of another edge.
  \item We describe implementations of a dynamic call graph and dynamic flow trace analysis of function values for JavaScript, both of which handle several hard-to-analyze JavaScript features.
  \item We present results and key observations from applying our techniques for the ACG algorithm~\cite{Feldthaus2013} and a suite of framework-based web applications.
\end{itemize}

The remainder of this paper is organized as follows.  \Cref{sec:background} provides background, and \Cref{sec:dynamic-analyses} describes our dynamic analyses.  \Cref{sec:missing-flows} presents our technique for automatically discovering root causes for missing edges.  \Cref{sec:implementation} gives details of our implementation.  \Cref{sec:study-setup} describes the setup of our study, and \Cref{sec:results} presents our results. \Cref{sec:related} discusses related work, and \Cref{sec:conclusions} concludes.

\section{Background}\label{sec:background}

We first give some background on challenges for JavaScript static analysis and on call graph construction.

\subsection{JavaScript analysis challenges}\label{sec:js-difficulties}

JavaScript programs often pose particularly difficult challenges for static analysis.  JavaScript includes numerous dynamic and reflective language features that are difficult to analyze, and unfortunately these features are used often in practice~\cite{Richards2010}.  We briefly present such features here; see previous work for detailed discussions (e.g.,~\cite{Richards2010,Sridharan2012,JensenJM12,ParkLR13}).  Tricky features include:
\begin{itemize}
  \item \textbf{Dynamic Property Accesses:} JavaScript object fields, or \emph{properties}, can be accessed using the syntactic form \code{x[e]}, where \code{e} is an arbitrary expression evaluating to a string property name.  Determining what memory locations may be accessed by an expression \code{x[e]} (fundamental to tracking data flow) can be a significant analysis challenge.  Further, if \code{e} evaluates to a property name that does not exist on \code{x}, a write to \code{x[e]} \emph{creates} the property rather than failing, making precise analysis even more challenging.
  \item \textbf{Eval:} JavaScript allows for evaluating arbitrary strings as code at runtime, most commonly via its \code{eval} construct or the \code{Function} constructor.  This dynamically-evaluated code is known to pose significant problems for static analysis~\cite{Richards2011,JensenJM12}.
  \item \textbf{With:} The \code{with} construct enables adding arbitrary variable bindings with a dynamically-constructed map~\cite{mdnwithdoc}.  As with \code{eval}, \code{with} usage complicates static analysis~\cite{ParkLR13}.
  \item \textbf{Getters and Setters:} A JavaScript property may be defined such that accessing the property actually invokes a \emph{getter} or \emph{setter} method with custom logic~\cite{gettersandsetters}. This feature makes it difficult to precisely identify the program locations where a function call can occur.
  \item \textbf{Reflective Calls:} JavaScript provides reflective methods to pass function parameters in flexible ways, e.g., binding the \code{this} parameter explicitly or passing arguments in an array~\cite{mdnfunctiondoc}.  Also, any function may read its formal parameters via a special \code{arguments} array, enabling variadic functions.  Finally, any function may be legally invoked with \emph{any} number of parameters, independent of how many formal parameters it declares.  Together, these features complicate tracking of inter-procedural data flow.
  \item \textbf{Native Methods:} JavaScript and the web platform provide a large standard library whose implementation is typically opaque to static analysis; hence, models must be constructed for a large number of these ``native'' methods.
\end{itemize}

While these root causes of difficult analysis are well known, our techniques enable measurement of their \emph{relative} impact on call graph recall for a set of target benchmarks.

\subsection{Call graph construction}\label{sec:cg-construction}

In a static call graph, nodes represent program methods, and an edge from $a$ to $b$ means that $a$ may invoke $b$ at runtime.\footnote{The call graph also includes information on which instruction in $a$, or \emph{call site}, may invoke $b$.}  The utility of a computed call graph $CG$ can be measured in terms of \emph{precision} and \emph{recall}.  Precision measures the number of infeasible edges in $CG$ (edges for calls that cannot occur in any execution), while recall measures the number of feasible call edges (those that \emph{can} occur in some execution) missing from $CG$.  Recall will be $100\%$ for any sound call graph construction technique, but as noted in \Cref{sec:intro}, many practical techniques sacrifice soundness for improved scalability and precision.  It is undecidable to compute the ``ground truth'' of possible calls for an arbitrary program, required to measure precision and recall perfectly.  Our evaluation (and previous work~\cite{Feldthaus2013,salis21pycg,Sui2020,nielsen21modular}) proceeds by exercising benchmarks using a best-effort process and then studying recall using the measured dynamic behaviors.

\tightpara{Static Flow Graphs} Our technique also relies on obtaining a \emph{static flow graph} from the static call graph analysis, to determine what dynamic data flow of function values was missed by the static analysis (see \Cref{fig:methodology-overview} and further discussion in \Cref{sec:missing-flows}).  In a flow graph, each node represents either a memory location (variables, object properties, etc.), a function value, or a call sites.  Edges in the flow graph are defined as follows: if the call graph analysis determines that a function value may be read from (abstract) memory location $m_1$ and then written to location $m_2$ (i.e., it may be directly copied from $m_1$ to $m_2$), the static flow graph should include an edge from $m_1$ to $m_2$.  So, flow graph edges should capture observed assignments of function values into variables and object properties, and passing of function values as parameters or return values to capture inter-procedural data flow.  Additionally, for a call $m_i(...)$, the flow graph should contain an edge from $m_i$ to a ``callee'' node for the call site (see example below).  With this construction, the static call graph should have an edge from call site $s$ to function $f$ iff there is a path from $f$ to the callee node for $s$ in the flow graph.

Graph representations are standard in analyses that track data flow~\cite{sridharan13alias}.  Further, any realistic JavaScript call graph construction algorithm must track function data flow, as JavaScript provides no basis for a cheaper technique (functions cannot be coarsely matched to possible call sites using types or even function arity).  Hence, we expect extraction of flow graphs from JavaScript call graph analyses will be straightforward.

\tightpara{Example} \Cref{fig:small-code-ex} gives a small running example for illustrative purposes.  \Cref{li:obj-create} creates an object with two fields \code{MyName} and \code{MyPhone}, respectively holding functions \code{f1} and \code{f2}.  \Cref{li:static-prop-call} reads and invokes \code{f1} using a \emph{static} property access (the property name is syntactically evident), whereas \cref{li:dyn-prop-call} reads and invokes \code{f2} using a dynamic property access.

\Cref{fig:missing-flow} shows the flow graph constructed by a variant of the call graph builder we study~\cite{Feldthaus2013} for the \Cref{fig:small-code-ex} example.  Edges represent the possible flow of function \code{f1} to the variable \code{v1}, then the object property \code{MyName}, and finally the call at \cref{li:static-prop-call}.  Given this path, the static call graph includes an edge from \code{main} to \code{f1}.  In contrast, the edge from the \code{MyPhone} property node to the call on \cref{li:dyn-prop-call} is missing in \Cref{fig:missing-flow}, due to the dynamic property access.  Our approach can determine that this missing flow graph edge leads to a missing \code{main}-to-\code{f2} edge in the call graph, and further reason that a dynamic property access is the root cause of the missed edge.

\begin{figure}
\begin{lstlisting}
function main() {
  var v1 = function f1() { return "John"; } /*@ \label{li:create-f1} @*/
  var v2 = function f2() { return "555-1234"; } /*@ \label{li:create-f2} @*/
  var obj = { MyName: v1, MyPhone: v2 }; /*@ \label{li:obj-create} @*/
  obj.MyName(); /*@ \label{li:static-prop-call} @*/
  obj["My" + "Phone"](); /*@ \label{li:dyn-prop-call} @*/
}
main();
\end{lstlisting}
\caption{Small example to illustrate our techniques.}
\label{fig:small-code-ex}
\end{figure}
\begin{figure}
\centering
\includegraphics[width=0.95\linewidth,height=3.5cm,keepaspectratio]{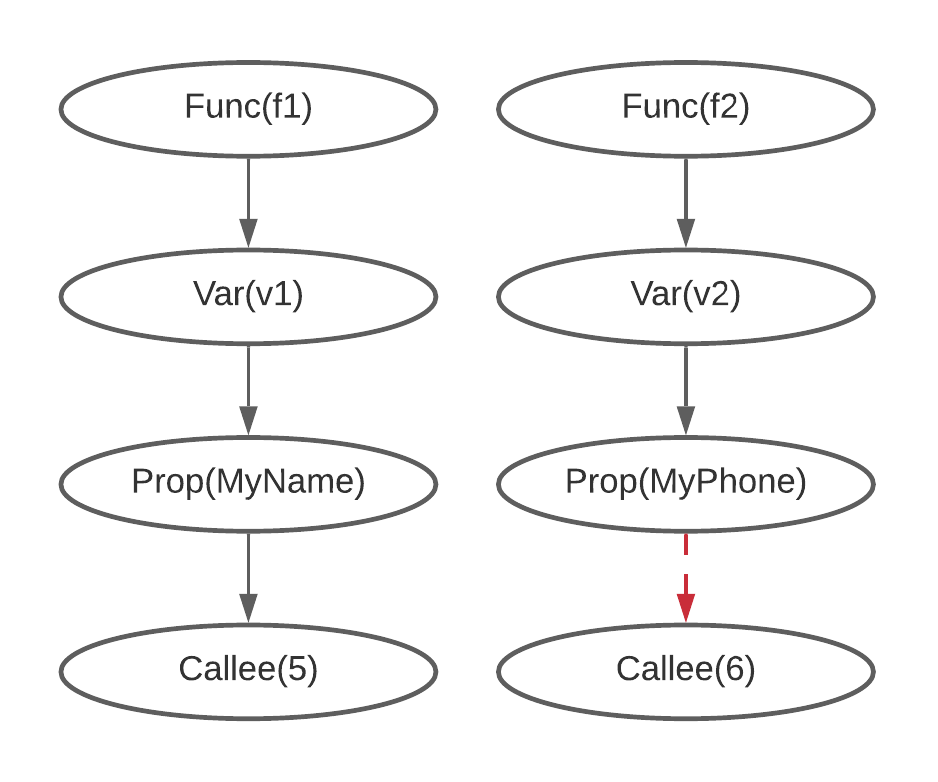}
\caption{Flow graph for \Cref{fig:small-code-ex}.  The red dashed edge is missing from the graph.}
\label{fig:missing-flow}
\end{figure}

\section{Dynamic Analyses}\label{sec:dynamic-analyses}

Our technique uses dynamic analyses to determine calls and data flows of function values occurring in executions of a program; this information is then compared with that in the static call graph and flow graph to detect missing flows (see \Cref{sec:missing-flows}).  Here we describe the dynamic analyses at a high level; we discuss implementation challenges related to complex JavaScript language constructs (such as those listed in \Cref{sec:js-difficulties}) in \Cref{sec:implementation}.

\tightpara{Dynamic Call Graphs} A dynamic call graph captures the calls that occurred in an execution (or set of executions) of a program.  As with static call graphs, nodes represent program methods and edges represent invocations between methods. At a high level, constructing dynamic call graphs only requires recording the actual functions invoked at each call instruction in some suitable data structure, and this type of analysis has been built many times before, including for JavaScript~\cite{HerczegL19}.  However, our analysis goes further by exposing call-related behaviors of some of the tricky JavaScript constructs outlined in \Cref{sec:js-difficulties}, crucial for a more complete understanding of static call graph recall.

\tightpara{Dynamic Flow Traces} Beyond dynamic call graphs, our technique requires \emph{dynamic flow traces} to find gaps in the data flow reasoning of static call graph builders.  A dynamic flow trace logs all data flow and invocation operations performed on function values.  The trace includes an entry for each creation of a function value (e.g., an expression \code{function () \{ ... \}}) and for each function call.  It also includes an entry for each read or write of a function value to or from a variable or object property.

As an example, here is an excerpt of the dynamic flow trace for the code in \Cref{fig:small-code-ex} (some details elided):
\begin{quote}
  \begin{footnotesize}
    \Call{Create}{\codify{f1,2}}; \Call{VarWrite}{\codify{v1,f1,2}};\\
  \Call{Create}{\codify{f2,3}}; \Call{VarWrite}{\codify{v2,f2,3}};\\
  \Call{VarRead}{\codify{v1,f1,4}}; \Call{PropWrite}{\codify{MyName,f1,4}};\\
  \Call{VarRead}{\codify{v2,f2,4}}; \Call{PropWrite}{\codify{MyPhone,f2,4}};\\
  \Call{PropRead}{\codify{MyName,f1,5}}; \Call{Invoke}{\codify{f1,5}};\\
  \Call{PropRead}{\codify{MyPhone,f2,6}}; \Call{Invoke}{\codify{f2,6}};\\
  \vspace{-\baselineskip}
\end{footnotesize}
\end{quote}
Each entry includes information on the function value being accessed and the location of the access (here, line numbers).  For property accesses, our traces only record the accessed property name, as the call graph techniques we studied in our evaluation do not distinguish base objects of accesses.  The trace could easily be extended to include base object identifiers if needed to study other analyses.

For handling of higher-order functions, the trace includes entries for parameter passing and returns of function values.  A call passing a function as a parameter is treated as a ``write'' of a parameter variable, which can be read via the formal parameter in the callee.  For returns, a \code{return} statement ``writes'' a special variable associated with the function's return value, which is ``read'' at the corresponding call site.

\section{Missing Flow Detection}\label{sec:missing-flows}

In this section, we describe our technique for discovering the \emph{missing flows} explaining why a static call graph is missing an observed dynamic call graph edge.  See \Cref{fig:methodology-overview} for our overall architecture.  Given a dynamic flow trace for a program, we first post-process the trace to discover the relevant \emph{dynamic copies} for a function call (\Cref{sec:dynamic-copies}).  Then, our technique matches these dynamic copies to the static flow graph, and automatically computes the missing flows relevant to each missing call edge (\Cref{sec:flow-graph-matching}).

\subsection{Finding Relevant Dynamic Copies for a Call}\label{sec:dynamic-copies}

Given a dynamic flow trace and an invocation of function $f$ at a call site, our technique computes the \emph{dynamic copies} by which $f$ was invoked at the site.  Dynamic copies capture data flow of function values at runtime---they are the dynamic analogue of the possible data flow captured in a static flow graph (\Cref{sec:cg-construction}).  A dynamic copy captures one of three operations on function values: (1) the value is \emph{created} and then stored in some memory location; (2) the value is \emph{copied} from one memory location to another; and (3) the value is read from a location and \emph{invoked}.  By computing the relevant dynamic copies for a particular call, our technique can expose which data flows may have been missed by the static analysis.

\begin{algorithm}[t]
  \caption{Finding dynamic copies for a call.}
  \label{alg:find-copies}
  \begin{small}
  \begin{algorithmic}[1]
  \Procedure{FindDynamicCopies}{$T$, $t_c$}
  \State $f \gets \textrm{function invoked by}\ t_c$ 
  \State $t_r \gets \textsc{PrecedingReadOrCreate}(T,t_c,f)$ \label{li:alg-initial-read}
  \State $C \gets \lbrack (t_r \xrightarrow{\mathit{invoke}} t_c) \rbrack$
  \While{$t_r$ is not a \textsc{Create} operation} \label{li:alg-loop-start}
    \State $t_w \gets \textsc{MatchingWrite}(T,t_r,f)$ \label{li:alg-find-matching-write}
    \State $t_{r'} \gets \textsc{PrecedingReadOrCreate}(T,t_w,f)$ \label{li:alg-find-next-read}
    \State $C \gets (t_{r'} \xrightarrow{t_w} t_r)::C$ \label{li:alg-add-dyn-copy}
    \State $t_r \gets t_{r'}$
  \EndWhile \label{li:alg-loop-end}
  \State \Return $C$
  \EndProcedure
  \Procedure{MatchingWrite}{$T$, $t_r$, $f$}
  \If{$t_r$ reads variable $x$}
  \State \Return $\textsc{PrecedingVarWrite}(T, t_r, f, x)$
  \ElsIf{$t_r$ reads property $\mathit{prop}$}
  \State \Return $\textsc{PrecedingPropWrite}(T, t_r, f, \mathit{prop})$
  \ElsIf{$t_r$ reads formal $p$ of function $f'$}
  \State // \textit{preceding invoke of $f'$ passing $f$ to $p$}
  \State \Return $\textsc{PrecedingInvoke}(T, t_r, f', f, p)$
  \ElsIf{$t_r$ is return value of call to $f'$}
  \State // \textit{preceding return of $f$ from $f'$}
  \State \Return $\textsc{PrecedingReturn}(T, t_r, f', f)$
  \EndIf
  \EndProcedure
  \end{algorithmic}
\end{small}
\end{algorithm}

Pseudocode for finding relevant dynamic copies appears in \Cref{alg:find-copies}.  We use sub-scripted $t$ variables for trace entries.  Given an entry $t_c$ for a call invoking function $f$ in trace $T$, \textsc{FindDynamicCopies} computes a list $C$ of the relevant dynamic copies, starting at the creation of $f$ and ending at the call.  Each dynamic copy is represented in the form $t_{r'} \xrightarrow{t_w} t_r$, read as: the function was read from memory by $t_{r'}$, and then copied to the memory location read by $t_r$, via write $t_w$.  The algorithm proceeds \emph{backwards} through the trace, starting at $t_c$ and reconstructing step-by-step how $f$ was copied through memory to reach the call site.

\Cref{alg:find-copies} first finds the read or create operation $t_r$ for $f$ most closely preceding $t_c$ in the trace (\cref{li:alg-initial-read}), corresponding to evaluation of $e$ in an invocation $e(...)$.\footnote{In certain corner cases, the closest preceding operation may not be the correct one; we discuss further under Limitations.}  $C$ is then initialized with $t_r \xrightarrow{\mathit{invoke}} t_c$, with the $\mathit{invoke}$ label indicating this is not a true copy, but instead the invocation of $f$.  

The loop at lines~\ref{li:alg-loop-start}--\ref{li:alg-loop-end} discovers relevant dynamic copies by matching writes and reads backward in the trace.  First, \Cref{li:alg-find-matching-write} finds the closest-preceding write operation $t_w$ that updated $t_r$'s location, using the \textsc{MatchingWrite} procedure.  \textsc{MatchingWrite}'s logic proceeds in cases, handing variables, object properties, formal parameters, and return values in turn.  For a read of property $\mathit{prop}$, the pseudocode matches with the most recent write to $\mathit{prop}$ on any object, matching the heap abstraction used by the call graph builder variants we study (see \Cref{sec:acg-details}).  For more precise call graph algorithms, the logic could easily be updated to also match the exact base object used in the property read operation.
Once the matching write $t_w$ is discovered, \cref{li:alg-find-next-read} finds the closest-preceding read or create $t_{r'}$, which ``produced'' $f$ for the write, and prepends a dynamic copy $t_{r'} \xrightarrow{t_w} t_r$ to $C$.

As an example, consider the call to \code{f2} on \cref{li:dyn-prop-call} in \Cref{fig:small-code-ex}.  Here are the relevant trace entries for that call visited by \Cref{alg:find-copies}:
\begin{quote}\begin{small}
  \Call{Create}{\codify{f2,3}};
  \Call{VarWrite}{\codify{v2,f2,3}};\\
  \Call{VarRead}{\codify{v2,f2,4}};
  \Call{PropWrite}{\codify{MyPhone,f2,4}};\\
  \Call{PropRead}{\codify{MyPhone,f2,6}};
  \Call{Invoke}{\codify{f2,6}}; \\
  \vspace{-\baselineskip}
\end{small}\end{quote}
Starting from the \textsc{Invoke} entry, the closest preceding read of \code{f2} is the \textsc{PropRead} of \codify{MyPhone} on \cref{li:dyn-prop-call}.  So, $C$ is initialized with $\Call{PropRead}{\codify{MyPhone,f2,6}} \xrightarrow{\mathit{invoke}} \Call{Invoke}{\codify{f2,6}}$.  The matching \textsc{PropWrite} for the read occurs on \cref{li:obj-create}, and its closest preceding read of \code{f2} is the \textsc{VarRead} on \cref{li:obj-create}.  Hence, we prepend a dynamic copy $\Call{VarRead}{\codify{v2,f2,4}}  \xrightarrow{t_{w_1}} \Call{PropRead}{\codify{MyPhone,f2,6}}$, where $t_{w_1} = \Call{PropWrite}{\codify{MyPhone,f2,4}}$.  Proceeding similarly, we reach the creation point of \code{f2} on \cref{li:create-f2}, prepend a dynamic copy $\Call{Create}{\codify{f2,3}} \xrightarrow{t_{w_2}} \Call{VarRead}{\codify{v2,f2,4}}$, where $t_{w_2} = \Call{VarWrite}{\codify{v2,f2,3}}$, and terminate.
    
\subsubsection*{Limitations} \Cref{alg:find-copies} assumes that the most-closely-preceding read of a function $f$ in the trace matches the subsequent write or invocation involving $f$.  In rare cases with parameter passing, this assumption may not hold, e.g.:
\begin{lstlisting}
function foo(p, q) { p(); }
function bar() {}
var x = bar;
var y = bar;
foo(x, y);
\end{lstlisting}
Assume we are trying to discover the dynamic copies for the call to \code{bar} on line 1.  Here is the relevant excerpt of the flow trace:
\begin{quote}\begin{small}
  ...; \Call{VarWrite}{\codify{x,bar,3}};
  \Call{VarWrite}{\codify{y,bar,4}};
  \Call{VarRead}{\codify{x,bar,5}}; \\
  \Call{VarRead}{\codify{y,bar,5}};
  \Call{Invoke}{\codify{foo,5}};
  \Call{VarRead}{\codify{p,bar,1}};
  \Call{Invoke}{\codify{bar,1}};
\end{small}\end{quote}
For the final \textsc{Invoke} of \code{bar}, the closest-preceding read is of formal parameter \code{p}.  The matching ``write'' is the \textsc{Invoke} of \code{foo} on line 5.  From here, the closest-preceding read of \code{bar} is from variable \code{y}, which is \emph{not} the parameter that gets passed in \code{p}'s position.  Hence, the analysis will discover an infeasible dynamic copy from the read of \code{y} to the read of \code{p}.  This simple case could be handled by using source locations during matching, but in cases involving recursion, dynamic call stacks would also need to be tracked.  We did not observe this behavior in any of our benchmarks, so we chose to employ the simpler technique of \Cref{alg:find-copies}.  

In some cases, the dynamic flow trace may be missing entries relevant to dynamic copies, due to JavaScript features like native methods and \code{with} (\Cref{sec:js-difficulties}) and also implementation limitations; see \Cref{sec:implementation} for details.  In such cases, our algorithm returns the subset of the relevant dynamic copies that it is able to reconstruct, and if possible notes a reason for its failure to find all copies.

\subsection{Flow Graph Matching}\label{sec:flow-graph-matching}

\newcommand{\fgsrc}{\ensuremath{\mathit{fgSrc}}\xspace}
\newcommand{\fgdst}{\ensuremath{\mathit{fgDst}}\xspace}

\begin{algorithm}[t]
  \caption{Finding missing flows in a flow graph for a call.}
  \label{alg:find-missing-flows}
  \begin{small}
  \begin{algorithmic}[1]
  \Procedure{FindMissingFlows}{$C$,$\mathit{CG}$,$\mathit{FG}$}
  \State $R \gets \emptyset$
  \For{each dynamic copy $t_{r'} \xrightarrow{t_w} t_r \in C$}
    \State $\fgsrc \gets \textsc{FlowGraphNode}(\mathit{FG},\mathit{t_{r'}})$ \label{li:fg-find-fgsrc}
    \State $\fgdst \gets \textsc{FlowGraphNode}(\mathit{FG},\mathit{t_r})$ \label{li:fg-find-fgdst}
    \If{$\fgsrc = \mathtt{null}$} \label{li:fg-fgsrc-null}
      \State $R \gets R \cup \textsf{MissingFGNode}(t_{r'})$
    \EndIf
    \If{$\fgdst = \mathtt{null}$}
      \State $R \gets R \cup \textsf{MissingFGNode}(t_r)$
    \EndIf \label{li:fg-fgdst-null-end}
    \If{$\fgsrc \neq \mathtt{null} \wedge \fgdst \neq \mathtt{null} \wedge \textsc{NoPath}(\mathit{FG},\fgsrc,\fgdst)$} \label{li:fg-check-path}
      \State $R \gets R \cup \textsf{MissingFGPath}(\fgsrc,\fgdst,t_{r'},t_w,t_r)$
    \EndIf
    \If{$t_{w}$ is a call} \label{li:fg-check-for-dep-call}
      \State $f \gets \textrm{function invoked by}\ t_{w}$ 
      \If{$\textsc{MissingFromCG}(\mathit{CG},t_{w},f)$} \label{li:fg-check-call-target}
        \State $R \gets R \cup \textsf{DependentCall}(t_w,f)$ \label{li:fg-add-dep-call}
      \EndIf
    \EndIf \label{li:fg-end-dep-call}
  \EndFor
  \State \Return $R$
  \EndProcedure
  \end{algorithmic}
\end{small}
\end{algorithm}

Given relevant dynamic copies for a call $c$ missed in the static call graph (discovered based on comparison with the dynamic call graph), we identify the missing flows for $c$ by matching the dynamic copies to the static flow graph extracted from the call graph builder.  (\Cref{sec:background} described static flow graphs, and \Cref{fig:missing-flow} gave an example.)  \Cref{alg:find-missing-flows} gives pseudocode for finding missing flows in a static flow graph.  The routine \textsc{FindMissingFlows} takes as inputs a list of dynamic copies $C$ produced by \textsc{FindDynamicCopies} in \Cref{alg:find-copies}, a static call graph $\mathit{CG}$, and the corresponding static flow graph $\mathit{FG}$.  Its result is a set of missing flows $R$, where each missing flow is one of three types: (1) \textsf{MissingFGNode}, indicating a node is missing in the flow graph, (2) \textsf{MissingFGPath}, indicating a path is missing in the flow graph, and (3) \textsf{DependentCall}, for when the absence of a flow is due to the absence of another call in the call graph.

For a dynamic copy $t_{r'} \xrightarrow{t_w} t_r$, the algorithm first tries to identify corresponding flow graph nodes \fgsrc and \fgdst (\cref{li:fg-find-fgsrc,li:fg-find-fgdst}).  In most cases, this matching is straightforward, done either by matching code entities or matching an accessed memory location to the flow graph node that abstracts it (we elide the details).  In some cases, the flow graph may not have a matching node, e.g., due to use of \code{eval} or due to an unmodelled property name from a dynamic property access.  In such cases, we record an \textsf{MissingFGNode} entry in the result (lines~\ref{li:fg-fgsrc-null}--\ref{li:fg-fgdst-null-end}).

If flow graph nodes $\fgsrc$ and $\fgdst$ are discovered, we next check for a \emph{path} from $\fgsrc$ to $\fgdst$ in the flow graph (\cref{li:fg-check-path}).  We must check for a path, rather than just an edge, since the static analysis may use temporary variables and assignments not present in the source code.  If no path is discovered, we note a \textsf{MissingFGPath} entry, retaining information about the dynamic copy to facilitate root cause labeling.

As an example, consider again the call to \code{f2} in \Cref{fig:small-code-ex}, and the corresponding dynamic copies described in \Cref{sec:dynamic-copies}.  In the \Cref{fig:missing-flow} flow graph for the code, there are matching nodes for all the copy locations, but there is no path matching the final copy $\Call{PropRead}{\codify{MyPhone,f2,6}} \xrightarrow{\mathit{invoke}} \Call{Invoke}{\codify{f2,6}}$.  So, the single missing flow computed for this case is a \textsf{MissingFGPath} entry with the details of this dynamic copy.  Given this information, a root cause labeler can discover that the flow was missed due to the dynamic property access; see \Cref{sec:root-cause-labeling}.

\tightpara{Dependent calls} Lines~\ref{li:fg-check-for-dep-call}--\ref{li:fg-end-dep-call} handle \emph{dependent calls}, where a path corresponding to a parameter passing or return dynamic copy is missing from the flow graph due to \emph{some other} missed call.  Consider this example:
\begin{lstlisting}
function f() { ... }
var x = { foo: function f2() { return f; } };
var y = x["fo"+"o"]();
y();
\end{lstlisting}
For the optimistic ACG call graph algorithm we use in our evaluation (see \Cref{sec:acg-details}), the calls to \code{f2} at line 3 and to \code{f} at line 4 will be missing in the call graph.  When finding missing flows for the line 4 call, a missing path for the function return dynamic copy at line 3 is discovered.  However, the issue with the analysis is not that it does not model returns of function values; this flow was missed \emph{because} the call target at line 3 was missed, so no flow could be discovered from the appropriate callee.  Our discovery of missing flows must account for such cases, to enable accurate quantification of root causes.

To handle dependent calls, \Cref{alg:find-missing-flows} checks at line~\ref{li:fg-check-for-dep-call} if the ``write'' operation for the copy was a call.  (Recall from \Cref{sec:dynamic-analyses} that calls are treated as the writes for parameter passing or function returns.)  If so, and if the static call graph is missing the relevant target for the call (line~\ref{li:fg-check-call-target}), we add a \textsf{DependentCall} missing flow to the result (line~\ref{li:fg-add-dep-call}).  

When counting the frequency of root causes, for dependent calls, we \emph{reuse} the root causes for one call as the root causes for the other.  For the example above, the dynamic property access at line 3 is identified as the single root cause for the missing calls at lines 3 and 4.  All results presented in \Cref{sec:results} precisely account for dependent calls.

\tightpara{Root Cause Labeling} Given a set of missing flows, quantification of root cause prevalence requires attributing a \emph{root cause label} to each missing flow.  The root cause labels may be specific to the call graph construction algorithm being studied, and must be developed with knowledge of the soundness gaps in the algorithm.  Additionally, root cause labeling may be performed with different levels of granularity, depending on what information is required by the analysis developer.  In \Cref{sec:root-cause-labeling}, we discuss the root cause labeling strategies used in our example study of the ACG call graph algorithm~\cite{Feldthaus2013}.

\section{Implementation}\label{sec:implementation}

\tightpara{Dynamic analyses} We implemented our dynamic call graph (DCG) and dynamic flow trace analyses (\Cref{sec:dynamic-analyses}) atop the Jalangi framework~\cite{SenKBG13a},\footnote{We use version 2 of Jalangi, available at \url{https://github.com/Samsung/jalangi2}.} which leverages source code instrumentation. While this instrumentation approach is more maintainable and portable than the alternatives, a downside is that the semantics of certain language constructs are not exposed in a straightforward way at the source level. In spite of source code instrumentation's limitations, one of its primary advantages is that it does not require modification of a JavaScript engine. Production JavaScript engines in browsers are challenging to modify, for two reasons: (1) they have complex implementations, so any change will require considerable engineering effort; and (2) they evolve rapidly, making it difficult to maintain an analysis. We use Jalangi2 to instrument JavaScript programs with our analysis code because it is easy to maintain and can work across different JavaScript engines. The tool allows us to perform analyses even when certain fragments of the source code are not instrumented. Our analyses contain significant extra logic to capture the behavior of several hard-to-analyze constructs as accurately as possible, despite the limitations of source instrumentation.

As an example, our DCG analysis exposes many callbacks from ``native'' library functions.  Such callbacks occurred regularly in the benchmarks used in our study, e.g., using \code{Function.prototype.call}, as shown in this small example:
\begin{lstlisting}
  function foo() { }
  foo.call(this);
\end{lstlisting}
Line 2 invokes \code{foo} via \code{call}, but Jalangi does not expose the invocation directly, as it cannot instrument \code{call}.  Instead, Jalangi exposes the invocation of \code{call}, followed by the start of execution in \code{foo}, but with no explicit invocation of \code{foo}.  To handle such cases, our DCG analysis maintains its own representation of the call stack.  Upon invocation of a native method, a marker for the method is pushed on the call stack.  Then, at the entry of a (non-native) method, if the top of our call stack is a native method marker, we record the fact that a native callback occurred.  For the above case, the dynamic call graph will include an invocation of the \code{call} native method at line 2, and also an invocation of \code{foo} from \code{call}, as desired.\footnote{Our technique does not yet precisely handle cases with multiple levels of native calls, such as \lstinline[basicstyle=\ttfamily\footnotesize]{Array.prototype.map.call(...)}; we plan to add further modeling for such cases in the future.\label{fn:multiple-native}}

Our DCG analysis also exposes getter and setter calls, and calls to and from dynamically-evaluated code.  For getters and setters, the analysis detects their presence via a library API~\cite{mdnpropertydescriptor}.  If a getter or setter is detected at a property access, it is treated as a call site and the call edge is recorded.  We leverage Jalangi's built-in support for dynamic code evaluation via \code{eval} or \code{new Function}; the relevant code string gets instrumented at runtime, so our analysis has visibility into calls into or out of such code.

Our dynamic flow trace analysis also includes special handling of some challenging JavaScript features.  The analysis distinguishes getters and setter calls using specially-marked \textsc{Invoke} entries, to enable tracking getter and setter use as a root cause.  For uses of the \code{arguments} array to access parameters, we generate relevant property write entries at a function entry as ``synthetic'' entries (not corresponding to explicit source code).  To handle \code{eval}-like constructs, any trace entry from the evaluated code includes a special source location marking it as from code executed via \code{eval}.

JavaScript has a very broad set of features and native methods requiring special handling, and our dynamic analyses still do not model all such features.  For the flow trace analysis, in certain cases a property write or read occurs in an unmodelled native method, and hence is missed in the trace.  The analysis generates special entries to model memory accesses performed by commonly-used library methods, such as \code{push} and \code{pop} on arrays.  We have not fully modeled all reflective constructs like \code{Object.defineProperty}~\cite{mdnobjdefineproperty}.
Also, use of the \code{with} construct can thwart our technique, as it is not fully supported by Jalangi.  (We note that all relevant uses of \code{with} in our benchmarks appeared \emph{within} an eval construct,\footnote{For example, see this code from the Knockout framework: \url{https://tinyurl.com/1jxtrpz3}} posing a severe challenge for static analysis.)

In terms of performance, we implemented some optimizations to reduce the size of the dynamic flow trace for larger benchmarks.  First, we limited tracing to only those function values that could be involved in a missing edge in the static call graph, based on the creation site of the function.  Second, we track a unique identifier for each function value using Jalangi's shadow memory functionality, and once the call site with the missing static call graph edge executes, we disable flow tracing for the corresponding value.

To generate dynamic call graphs and flow traces, we exercised our benchmarks manually and recorded the actions as Puppeteer~\cite{pptr.dev} automation scripts to allow for repeatable runs;  \Cref{sec:benchmarks-and-harness} details the coverage obtained for benchmarks in our study.

\tightpara{Missing Flow Detection} The missing flow detection algorithms of \Cref{sec:missing-flows} are implemented in \code{1154} lines of Python code.  For the most part, detecting missing flows in the static flow graph given a dynamic flow trace was straightforward.  Some effort was required to match source locations provided by WALA~\cite{wala} for JavaScript constructs (our use of WALA is detailed in \Cref{sec:acg-details}) with what was observed by the dynamic analyses.  In the process of ensuring this matching was precise, we contributed a couple of fixes to WALA, and also found and fixed a longstanding issue with incorrect source locations in the Rhino JavaScript parser~\cite{mozillarhino}.\footnote{\url{https://github.com/mozilla/rhino/pull/809}}

\section{Study Setup}\label{sec:study-setup}

Here, we detail the setup of our study of root causes of missed call graph edges for framework-based web applications.  We describe the ACG call graph algorithm used in our study (\Cref{sec:acg-details}), describe how we performed root cause labeling for this algorithm (\Cref{sec:root-cause-labeling}), and then present our benchmarks and how they were exercised (\Cref{sec:benchmarks-and-harness}).

We note that the main purpose of our study was to show the potential of our techniques to give useful insights on the relative importance of different root causes for missed static call graph edges.  We do \emph{not} claim that the results for the benchmarks used in our study will generalize to any broad class of framework-based web applications.  A study of a wider variety of benchmarks, to obtain generalizable insights on root causes across JavaScript applications, is beyond the scope of this work.

\subsection{The ACG algorithm}\label{sec:acg-details}

In our evaluation, we studied variants of the approximate call graph (ACG) algorithm of Feldthaus et al.~\cite{Feldthaus2013}.  The ACG algorithm was designed to entirely skip analysis of many challenging JavaScript language features, while still providing good precision and recall for real-world programs.  ACG leverages the insight that many dynamic property accesses in JavaScript are correlated~\cite{Sridharan2012}, with a paired dynamic read and write used to copy a property from one object to another.  By using a \emph{field-based} handling of object properties~\cite{Heintze2001} (treating each property as a global variable), ACG could ignore dynamic property accesses entirely and still provide good recall, assuming most accesses are correlated.  %

Feldthaus et al.~\cite{Feldthaus2013} describe \emph{pessimistic} and \emph{optimistic} variants of ACG, differing in their handling of inter-procedural flow.  Pessimistic ACG only tracks data flow across procedure boundaries in limited cases, whereas optimistic ACG performs full inter-procedural tracking.  We performed root cause quantification for both variants in our study.

Our study uses the open-source implementation of ACG in WALA~\cite{wala}.  This implementation directly builds a flow graph during call graph building, which we serialize alongside the computed call graph.  The WALA implementation also includes partial handling of the \code{call} and \code{apply} reflective constructs for parameter passing~\cite{mdnfunctiondoc}.  In the optimistic variant, interprocedural flow is handled fully for \code{call}, but only return values are handled for \code{apply} (as it passes parameters via arrays, which is hard to analyze).  We confirmed via inspection that the WALA implementation of ACG has no handling of getters and setters, \code{eval}, and \code{with}.  

\subsection{Root Cause Labeling}\label{sec:root-cause-labeling}

We implemented root cause labeling for missing flows based on the gaps we observed in the WALA implementation~\cite{wala} of the ACG algorithm~\cite{Feldthaus2013}.  For a different algorithm or implementation, some different root causes may be required, but we expect significant overlap, as several root causes pertain to challenging language features that many techniques handle unsoundly (e.g., \code{eval}).  The referenced root cause names are also used when discussing their prevalence in \Cref{sec:top-root-causes}.

For \textsf{MissingFGNode} (see \cref{sec:flow-graph-matching}), in some cases, there is no node representing the creation of a function value in the flow graph.  If the function was from the standard library, we assigned the label ``Call to unmodelled native function,'' as WALA was likely missing a model for the function.  In cases where the function was created via a call to \code{new Function} (unhandled by the ACG implementation), we assigned the label ``Creation via Function constructor.''

In other \textsf{MissingFGNode} cases, the node representing the call site itself is missing.  For this case, a common root cause label is ``Call to getter/setter,'' as getters and setters are not modeled by ACG.  Also, the ``Calls from unmodelled native functions'' label captures cases where an unmodeled native function calls back into application code.  Finally, for a dynamic property access, if the property name is never used as part of a non-dynamic property access, the flow graph may not have a node for the property, in which case we use the label ``Dynamic Property Access.''

For \textsf{MissingFGPath}, one possible root cause is ``Dynamic Property Access,'' which can be identified by the corresponding dynamic reads / writes.  For the pessimistic ACG variant, paths may be missing since the algorithm does not model passing function values as parameters or returning function values; we use the labels ``Parameter Pass'' and ``Function return'' for these scenarios.  For both ACG variants, the ``Parameter Pass'' label is also used to reflect passing of parameters in an array via \code{Function.prototype.apply}.

In the case of dynamically-evaluated code (the ``Use of Eval'' and ``Eval via new Function'' labels), many relevant nodes may be missing from the static flow graph.  We assign an appropriate root cause in these cases by recording in the flow trace which events occurred in dynamically-evaluated code (\Cref{sec:implementation}).  Note that we \emph{prioritize} the \code{eval}-related root causes over others; e.g., if there is a relevant dynamic property access in \code{eval}'d code, we will assign the \code{eval}-related root cause, even though it is possible the analysis also could not handle the property access.  We chose this labeling due to the high difficulty of handling \code{eval} constructs in static analysis; for an analysis with significant support for \code{eval} a different choice may be appropriate.

Finally, as noted in \Cref{sec:dynamic-copies}, in certain cases we cannot compute all dynamic copies for a call.  For these cases, our technique makes a base-effort attempt to assign an appropriate root cause label.
``Call to bounded function'' captures missing handling of the \code{Function.prototype.bind} feature~\cite{mdnfunctiondoc}.  The ``Multiple levels of native functionality'' label captures cases where native methods are invoked reflectively (see \Cref{fn:multiple-native}).    Finally, we identify the ``Use of With'' root cause by tracing objects used in \code{with} statements and identifying when an unmatched variable corresponds to a \code{with} object property.

As \Cref{sec:top-root-causes} will show, dynamic property accesses are the most frequent root cause of missing call graph edges for our benchmarks.  To further understand these root-cause accesses, we also implemented a finer-grained labeling for them, based on the expression used for the property name.  This more granular labeling is described in \Cref{sec:property-name-flow}.

\subsection{Benchmarks and Harness}\label{sec:benchmarks-and-harness}

\begin{table}
\setlength\tabcolsep{1pt}
\centering
\begin{tabular}{|C{5.5em}|C{4em}|C{5em}|C{5.5em}|C{5.5em}|}
\hline & Total LoC
& Application LoC & Framework/
Library LoC & Application Stmt. Coverage \\ 
\hline
AngularJs & 12091 & 256 & 11835 & 81.08\%\\ 
\hline
Backbone & 9003 & 216 & 8787 & 99.74\%\\ 
\hline
KnockoutJs & 1044 & 129 & 915 & 98.98\%\\ 
\hline
KnockbackJs & 15836 & 199 & 15637 & 99.73\%\\ 
\hline
CanJs & 11371 & 129 & 11242 & 100\%\\ 
\hline
React & 24855 & 383 & 24472 & 99.21\%\\ 
\hline
Mithril & 1433 & 252 & 1181 & 99.61\%\\ 
\hline
Vue & 7667 & 124 & 7543 & 97.73\%\\ 
\hline
VanillaJs & 751 & 561 & 190 & 98.10\%\\ 
\hline
jQuery & 9526 & 171 & 9355 & 99.59\%\\
\hline\hline
Juice Shop & >65000 & 15092 & >50000 & 36\%\\
\hline
\end{tabular}
\caption{Benchmark Statistics.}
    \label{tab:BenchmarkStats_table}
\end{table}

For benchmarks, our study used several programs from the TodoMVC suite~\cite{todomvc}.  TodoMVC contains many implementations of a simple web-based todo list application, with each implementation using a different web framework or language.  The suite is designed to help developers compare different model-view-controller (MVC) frameworks.
Because the suite contains idiomatic implementations of the same functionality across frameworks, it provides an opportunity to compare sources of missing call graph edges across frameworks.

To test with a larger web application, we also included OWASP Juice Shop~\cite{juiceshop}, an AngularJs-based program that is a standard benchmark for security analyses.  Counting the size of framework / library code for Juice Shop is difficult, as the code base does not clearly separate third-party code used as part of the web site from libraries used only to deploy the site; we conservatively estimated the framework / library code to be greater than 50 kLoC.

\Cref{tab:BenchmarkStats_table} gives statistics for our benchmarks.  The TodoMVC benchmarks are named based on the web framework that they use.  The TodoMVC applications range from 751--24,855 lines of code, with framework sizes varying widely.  We chose all eight of the JavaScript-framework-based implementations that worked with our infrastructure.\footnote{Some implementations used newer JavaScript language features not yet supported by Jalangi.}  We also chose VanillaJS, which does not use any framework,\footnote{All implementations use a common base JavaScript library, accounting for the library code in VanillaJS.} and jQuery, for comparison purposes.

To exercise the TodoMVC applications, we wrote a harness to cover as much application code as possible, and in the end our script achieved application code statement coverage of 97\% or higher for nearly all benchmarks.  We studied all uncovered code manually, and found that it was either dead code or could not be exercised in a single run of the application (e.g., for the AngularJs version, a small amount of code would only run if the app were used and then restarted in offline mode).

For Juice Shop, we were unable to exercise the application beyond fully completing its initial loading, explaining the significantly lower code coverage.  Our infrastructure ran into scalability issues for deeper runs of Juice Shop, which we hope to fully address in the near future.  Still, simply loading Juice Shop exercised a large amount of code (its flow trace was nearly 5 times larger than any fully-exercised TodoMVC benchmark), making a study of missed call edges for the loading portion of the execution interesting on its own.

In terms of running times for our tools, dynamic call graph and flow trace collection each took between 30 and 60 seconds for each TodoMVC benchmark, varying based on the amount of code executed; this overhead is comparable to previous Jalangi-based dynamic analyses~\cite{SenKBG13a}.  Missing flow detection (\Cref{sec:missing-flows}) took time proportional to the size of the flow trace, ranging from around half a second (for VanillaJS) to around 10 minutes (for React).  Overall running time for Juice Shop was much longer (more than an hour total) due to its size and the aformentioned scalability bottlenecks it exposed.  We expect the missing flow detection times could be reduced significantly with a more optimized implementation.

\section{Results}\label{sec:results}

In this section, we present results from performing root cause quantification for our benchmarks.  The results show that our quantification techniques can provide interesting insights into the relative prevalence of different root causes for missing call graph edges.  We first give recall measurements for our benchmarks using multiple metrics in \Cref{sec:recall-measurements}.  Then, we discuss the top root cause labels for missed call graph edges in \Cref{sec:top-root-causes} and insights gained from this data.  Finally, we discuss results from performing a finer-grained labeling of missing flows related to dynamic property accesses (the most prevalent root cause) in \Cref{sec:property-name-flow}. 

\subsection{Recall Measurements}\label{sec:recall-measurements}

We measured call graph recall for our benchmarks by comparing the ACG static call graphs with our collected dynamic call graphs.  We first describe our methodology, and then present results.  We also measured call graph precision for all benchmarks, but as our new techniques focus on root causes for low recall, we do not discuss the precision results here; they are presented in \iftoggle{extended}{\Cref{sec:precision-details}}{an extended version of the paper~\cite{extendedPaperVersion}}.

\tightpara{Methodology} We used three different metrics to measure recall, suited to different client scenarios:
\begin{itemize}
  \item \textbf{Call site targets:} the set of targets at each call site present in the dynamic call graph.  This metric was used in the original ACG paper~\cite{Feldthaus2013}.  Recall is computed for each call site, and then averaged across call sites to produce recall for a benchmark.  This metric is most relevant to clients like code navigation in an IDE.
  \item \textbf{Reachable nodes:} the set of reachable methods, where roots are the entrypoints in the dynamic call graph.  This metric has been used in previous work~\cite{Sui2020}, and is relevant to clients like dead-code elimination.
  \item \textbf{Reachable edges:} the set of call graph edges whose source method is present in the dynamic call graph.  This metric is most relevant to clients doing deep inter-procedural analysis like taint analysis~\cite{GuarnieriPTDTB11}.
\end{itemize}

Given our collected data, we studied the following research questions:
\begin{itemize}
  \item \textbf{RQ1:} How does recall vary across the three metrics?
  \item \textbf{RQ2:} How does recall vary across benchmarks?
\end{itemize}

\begin{figure}[t]
  \centering
  \begin{subfigure}[b]{0.7\linewidth}
      \includegraphics[width=\textwidth, keepaspectratio]{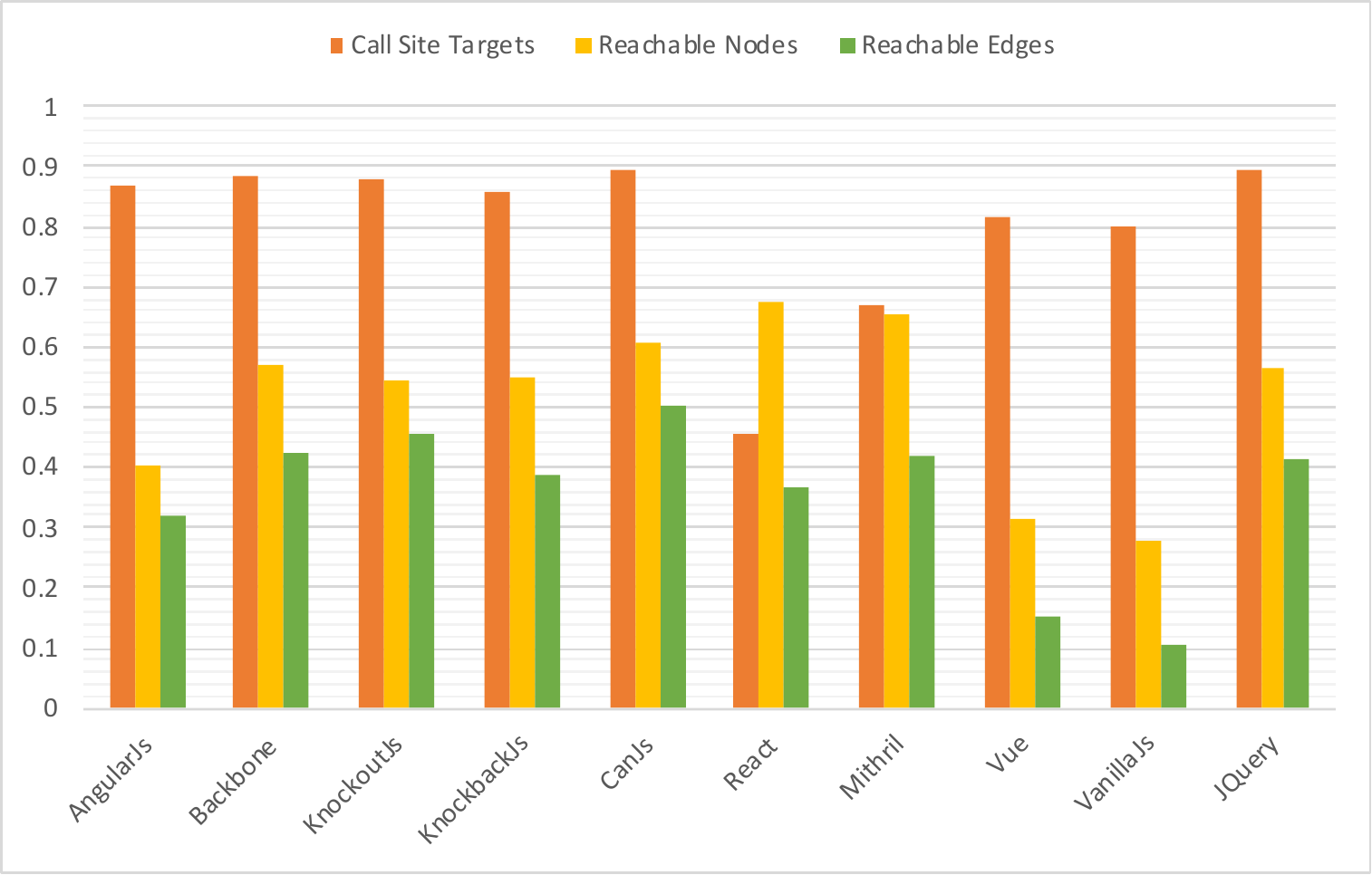}
      \caption{Pessimistic ACG.}
      \label{fig:pes-recall-bfr-imp}
  \end{subfigure}
  \begin{subfigure}[b]{0.7\linewidth}
      \includegraphics[width=\textwidth, keepaspectratio]{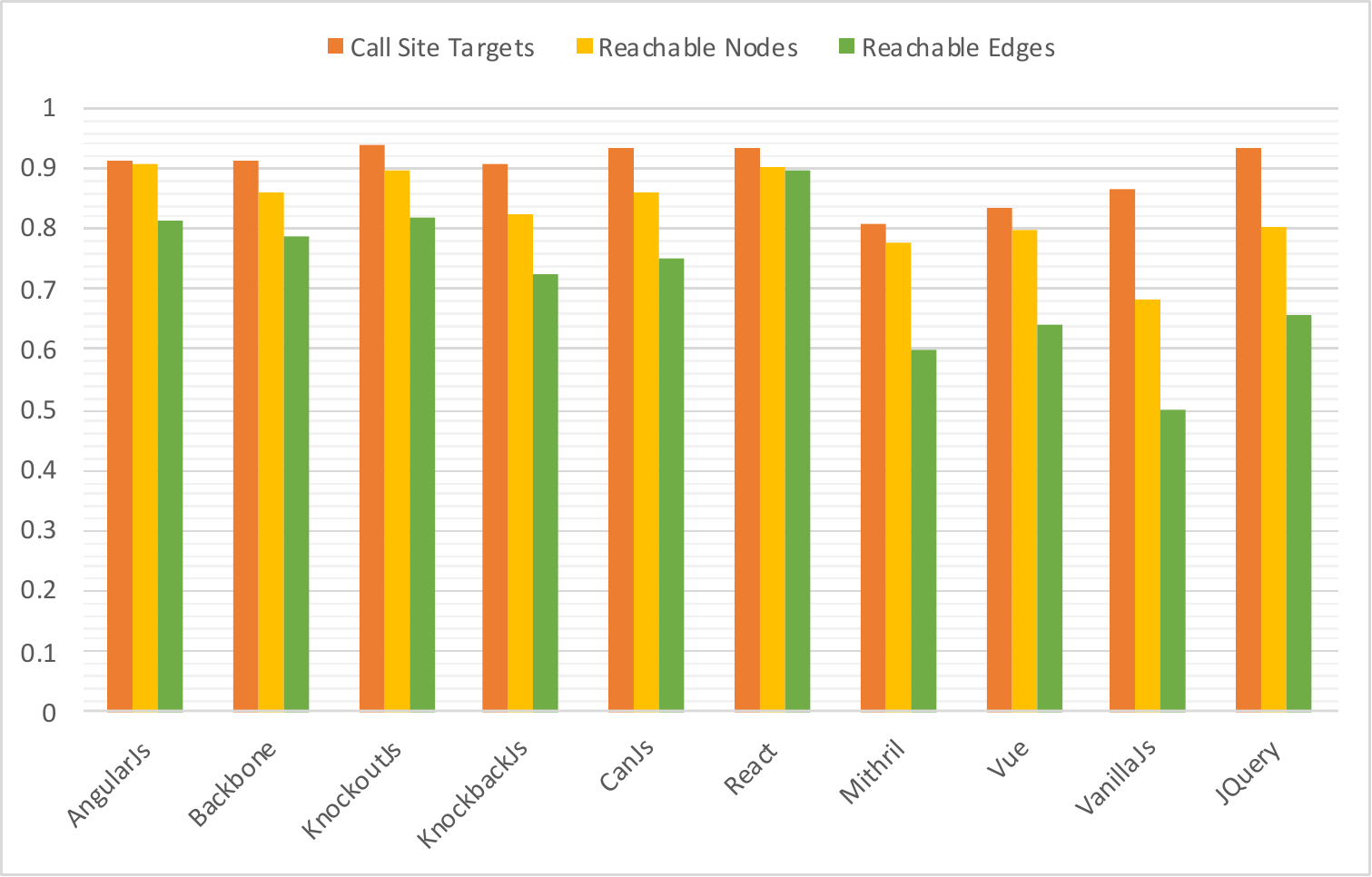}
      \caption{Optimistic ACG.}
      \label{fig:opt-recall-bfr-imp}
  \end{subfigure}
  \caption{Detailed recall results for our three metrics across the benchmarks.}
  \label{fig:recall-results-detailed}
\end{figure}

\begin{figure}[t]
  \centering
  \includegraphics[width=0.75\linewidth,keepaspectratio]{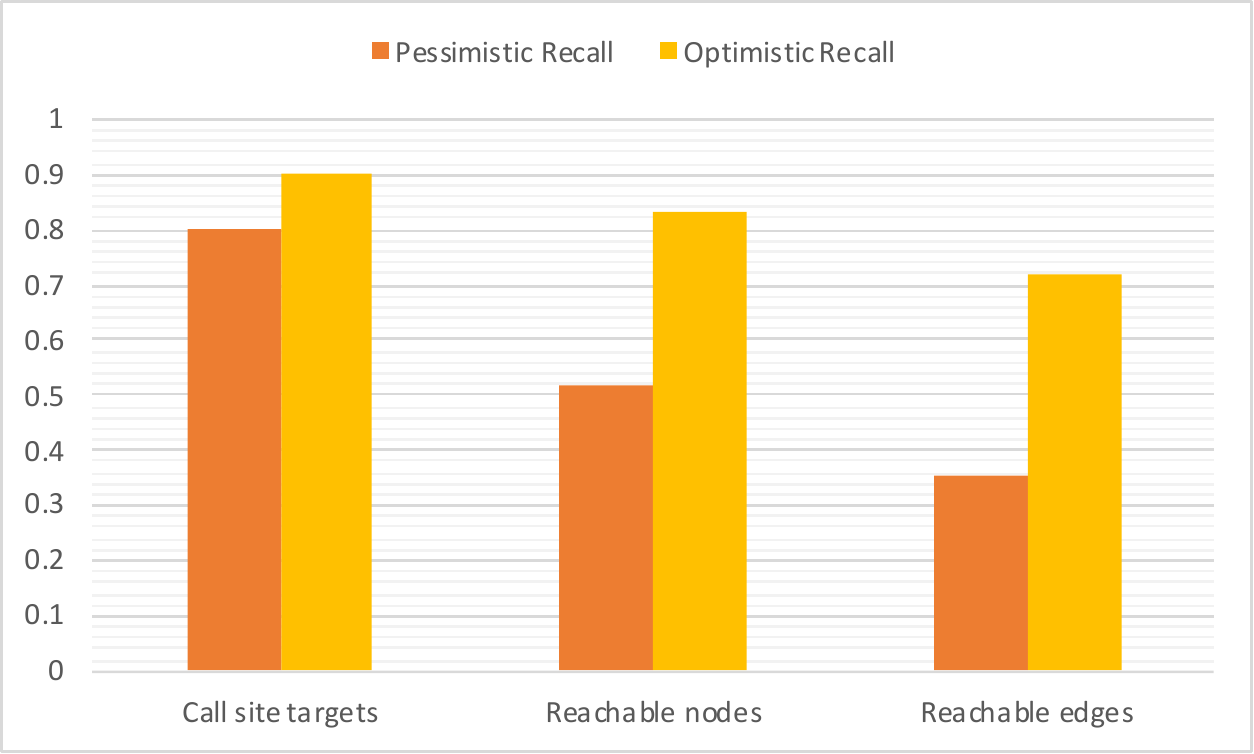}
  \caption{Average recall across benchmarks for original WALA ACG implementation.}
  \label{fig:avg-recall-detailed-b4}
\end{figure}

\tightpara{Results} \Cref{fig:recall-results-detailed} gives detailed recall results for WALA's original ACG implementation for each TodoMVC benchmark, with results for the pessimistic variant in \Cref{fig:pes-recall-bfr-imp} and for optimistic in \Cref{fig:opt-recall-bfr-imp}.  Average recall across the TodoMVC benchmarks is shown in \Cref{fig:avg-recall-detailed-b4}.

For RQ1, the data show that recall of ACG tends to suffer with more exacting metrics.  The ACG paper~\cite{Feldthaus2013} used the call site targets metric, and showed that both precision and recall were typically above 80\% for their benchmarks.  \Cref{fig:avg-recall-detailed-b4} shows that for our benchmarks, while recall is above 80\% for this metric for both the optimistic and pessimistic variants, recall decreases for the more exacting metrics, particularly for pessimistic analysis.

For RQ2, \Cref{fig:recall-results-detailed} shows that recall can vary widely across benchmarks.  In \Cref{sec:top-root-causes} we dig further into these differences, showing that root causes for low recall can also vary across the benchmarks.  For the TodoMVC React benchmark, recall is very high for the optimistic analysis but quite low for pessimistic.  In this case, the high recall for optimistic analysis comes at a cost of very low precision (less than 5\% for reachable edges; see \iftoggle{extended}{\Cref{sec:precision-details}}{the extended version of the paper~\cite{extendedPaperVersion}} for full details).  We suspect that some initial imprecision spirals out of control for optimistic analysis for React, leading to poor precision.  Previous work studied diagnosing imprecision root causes~\cite{LeePR20,AndreasenMN17,WeiTRD16}; such a study is out of scope here.  However, improving recall can lead to reduced precision, and this tradeoff must be minded when devising solutions to improving recall.

For Juice Shop, only the pessimistic ACG variant could run to completion; optimistic ACG could not complete within 64GB of memory.  Pessimistic ACG missed 15,060 edges that were present in the dynamic call graph.  Since our coverage for Juice Shop was significantly lower than the other benchmarks (see \cref{sec:benchmarks-and-harness}), we do not quantify the precision and recall of pessimistic ACG for the benchmark, nor do we include it in aggregate statistics.

\subsection{Root Cause Quantification}\label{sec:top-root-causes}

\begin{figure*}[htp!]
  \centering
  \begin{subfigure}[b]{0.48\linewidth}
      \centering
      \includegraphics[width=0.8\textwidth, keepaspectratio]{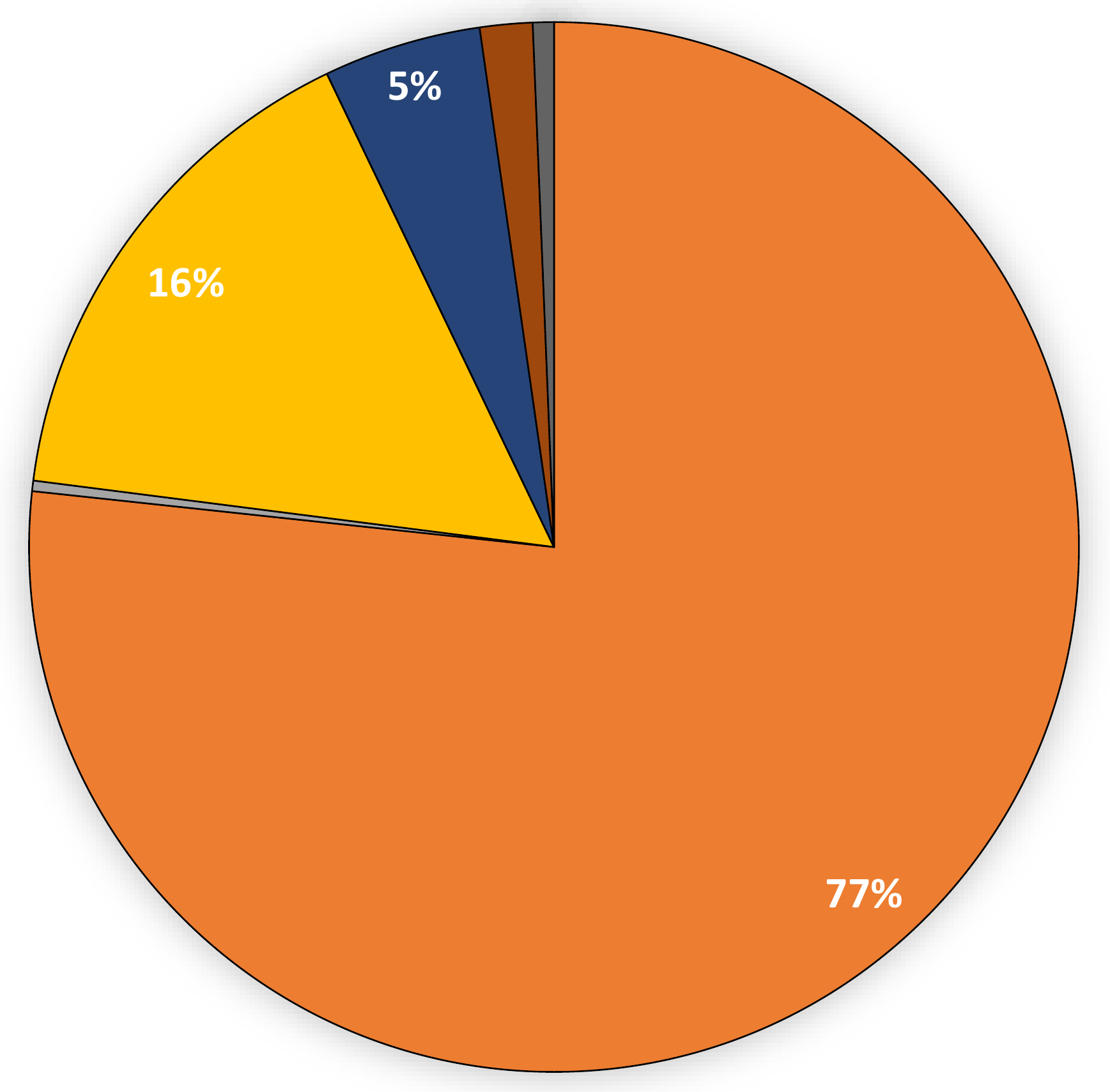}
      \caption{React}
      \label{fig:root-cause-react}
  \end{subfigure}%
~
  \begin{subfigure}[b]{0.48\linewidth}
      \centering
      \includegraphics[width=0.8\textwidth, keepaspectratio]{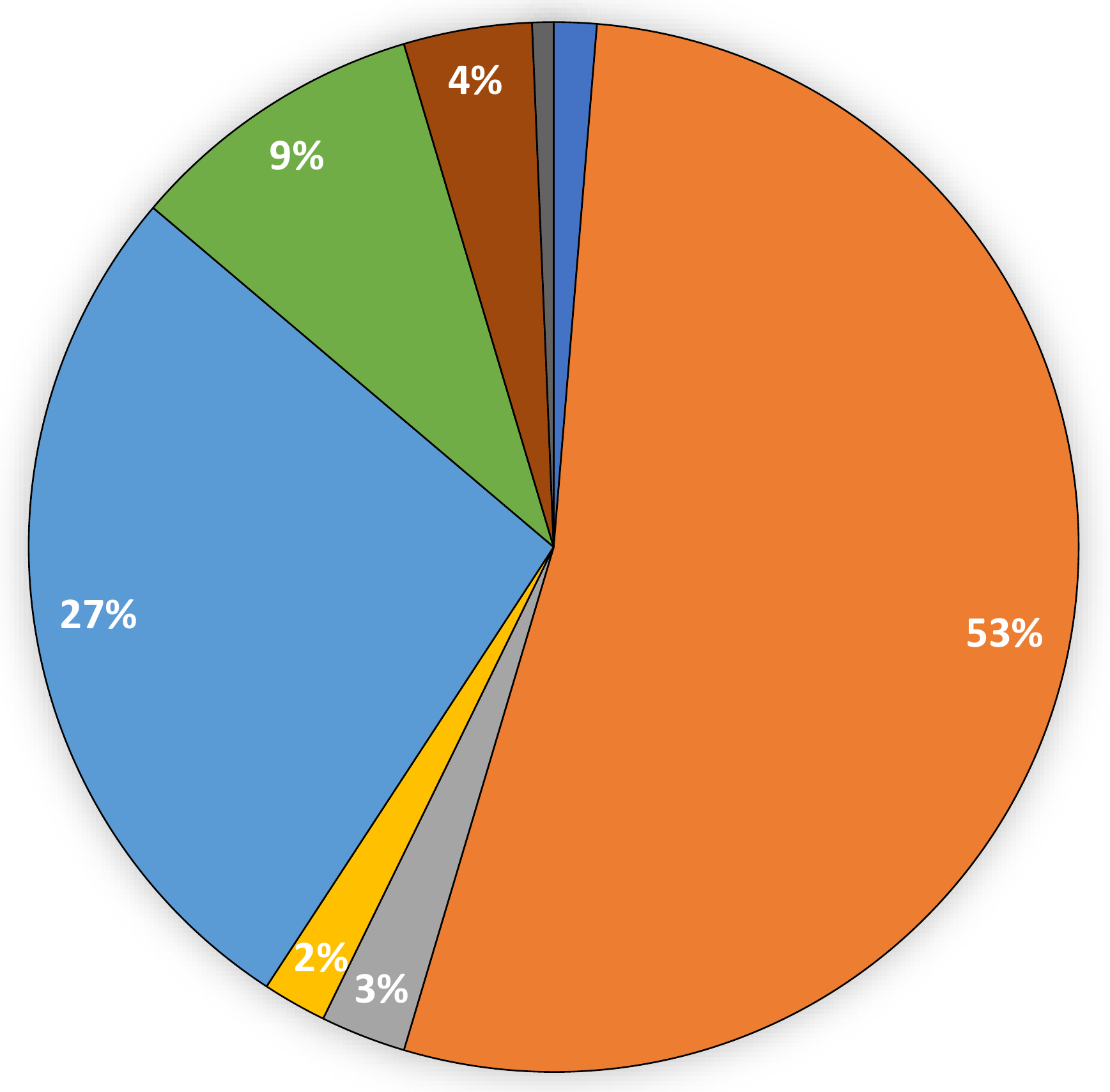}
      \caption{AngularJS}
      \label{fig:root-cause-angular}
  \end{subfigure}
~
  \begin{subfigure}[b]{0.48\linewidth}
      \centering
      \includegraphics[width=0.8\textwidth, keepaspectratio]{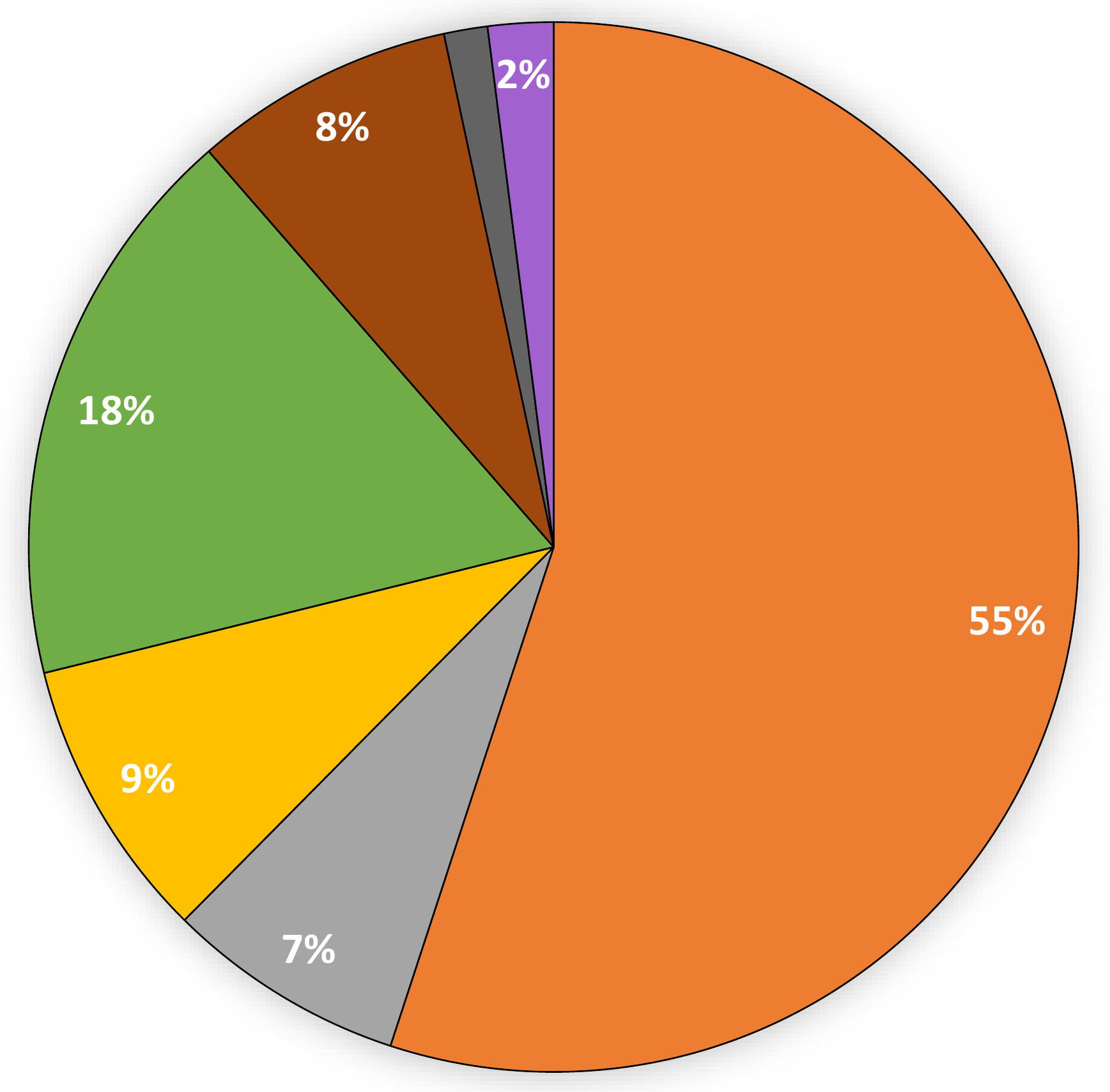}
      \caption{Vue}
      \label{fig:root-cause-vue}
  \end{subfigure}
\caption{Root causes for three TodoMVC benchmarks for optimistic ACG.}
\label{fig:root-cause-specific-benchmarks}
\end{figure*}

We present illustrative results from applying our techniques to quantify prevalence of root causes for missing call graph edges for our benchmarks.  Space does not allow a full presentation of all results; all experimental data is available in our artifact~\cite{acgArtifact}.  Here we focus on the following questions:
\begin{itemize}
  \item \textbf{RQ3:} What are the most common root causes for missed call graph edges?
  \item \textbf{RQ4:} Does the relative importance of root causes vary across benchmarks?
\end{itemize}
We compute root causes for each individual missed call edge in the static call graph, corresponding to the ``Reachable edges'' metric used to measure recall in \Cref{sec:recall-measurements}.  The color legend for the pie charts appears below \Cref{fig:root-cause-specific-benchmarks}.

\begin{figure}
\centering
\includegraphics[width=\linewidth,height=7cm,keepaspectratio]{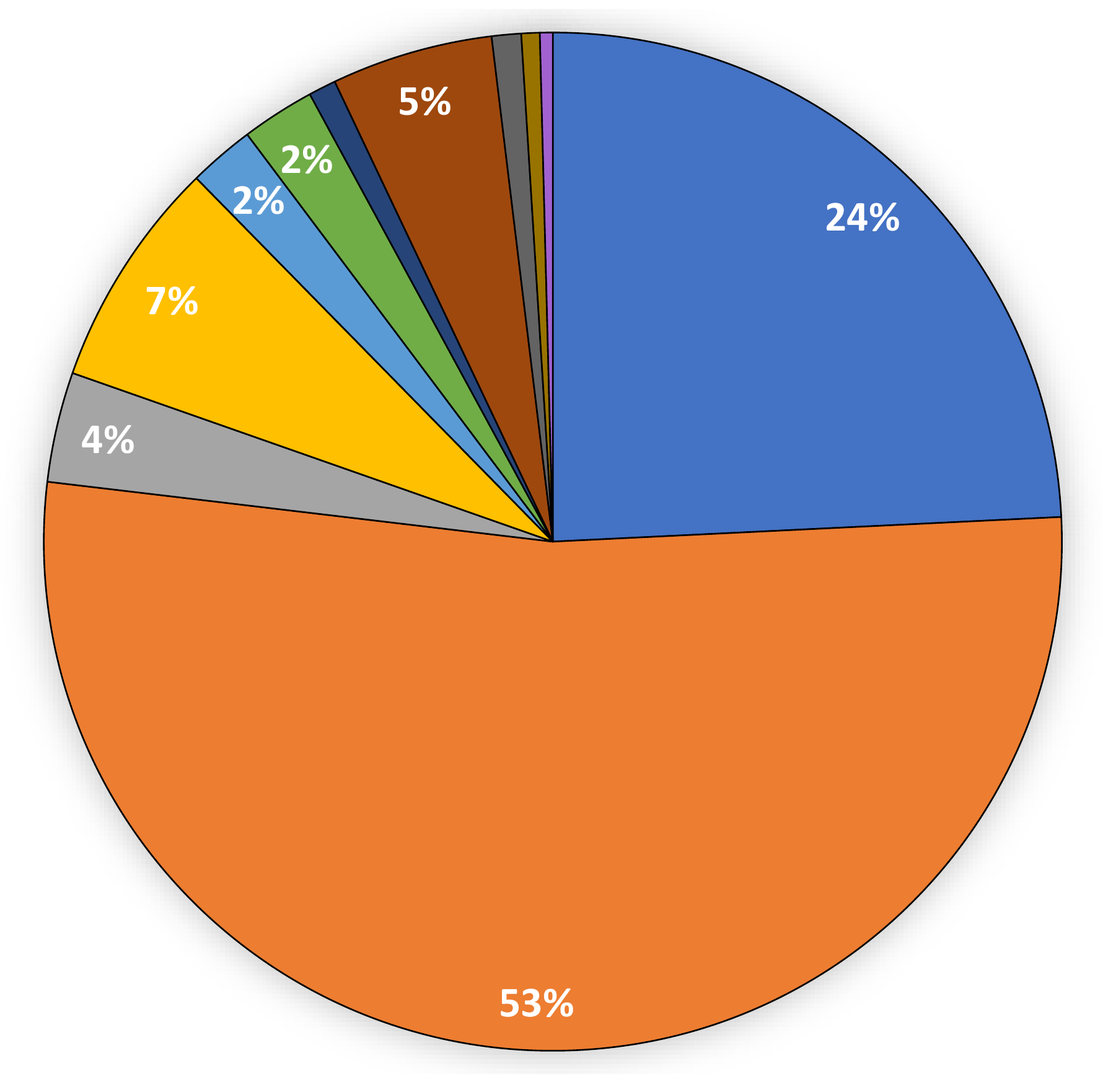}
\caption{Original root causes for optimistic ACG across TodoMVC, before WALA improvements.}
\label{fig:overall-todo-opt-old}
\end{figure}

\begin{figure*}[htp!]
  \centering
  \begin{subfigure}[b]{0.48\linewidth}
      \centering
      \includegraphics[width=0.8\textwidth,height=6cm,keepaspectratio]{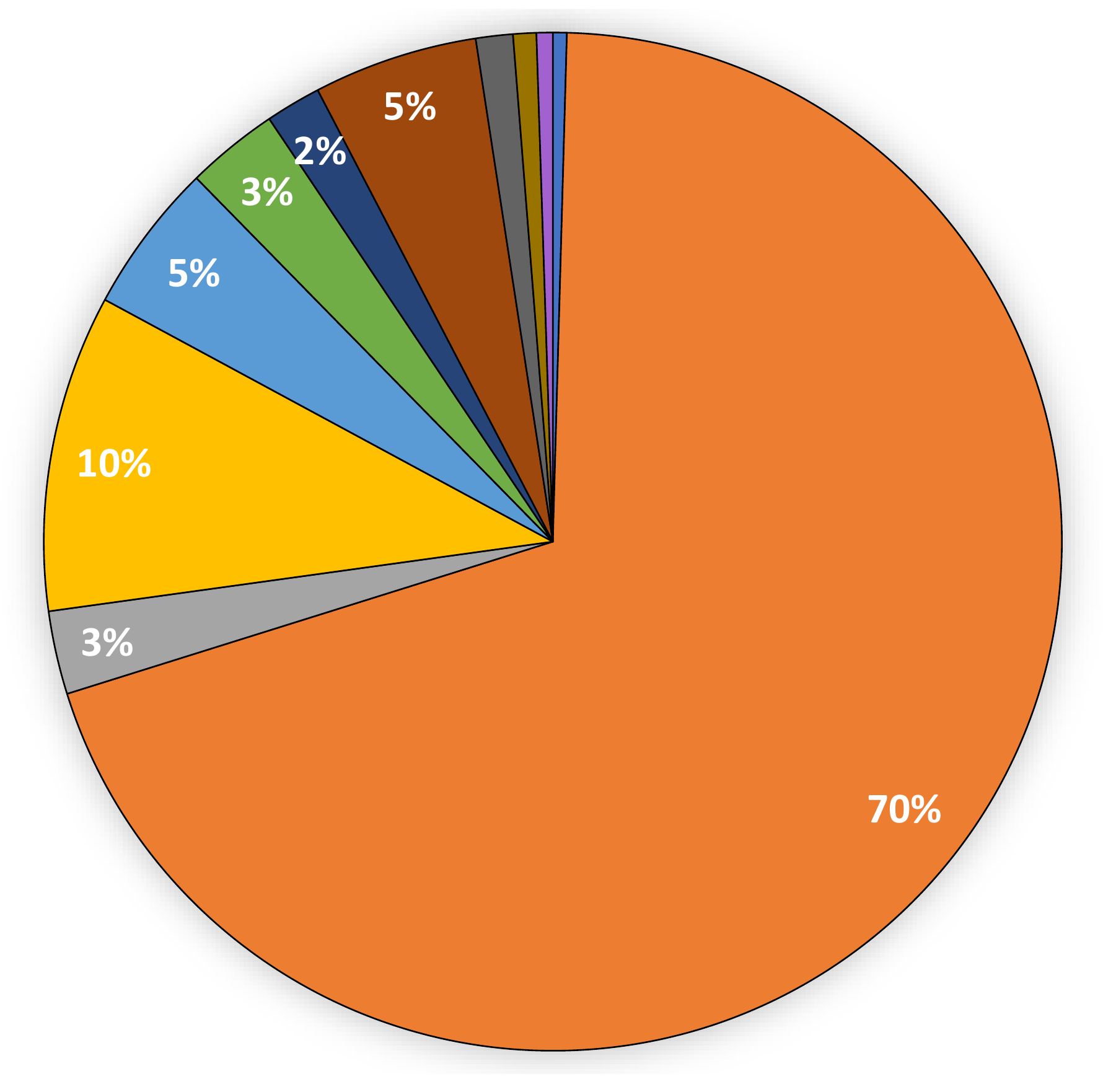}
\caption{Optimistic}
\label{fig:overall-todo-opt-new}
 \end{subfigure}%
~
  \begin{subfigure}[b]{0.48\linewidth}
      \centering
      \includegraphics[width=0.8\textwidth,height=6cm,keepaspectratio]{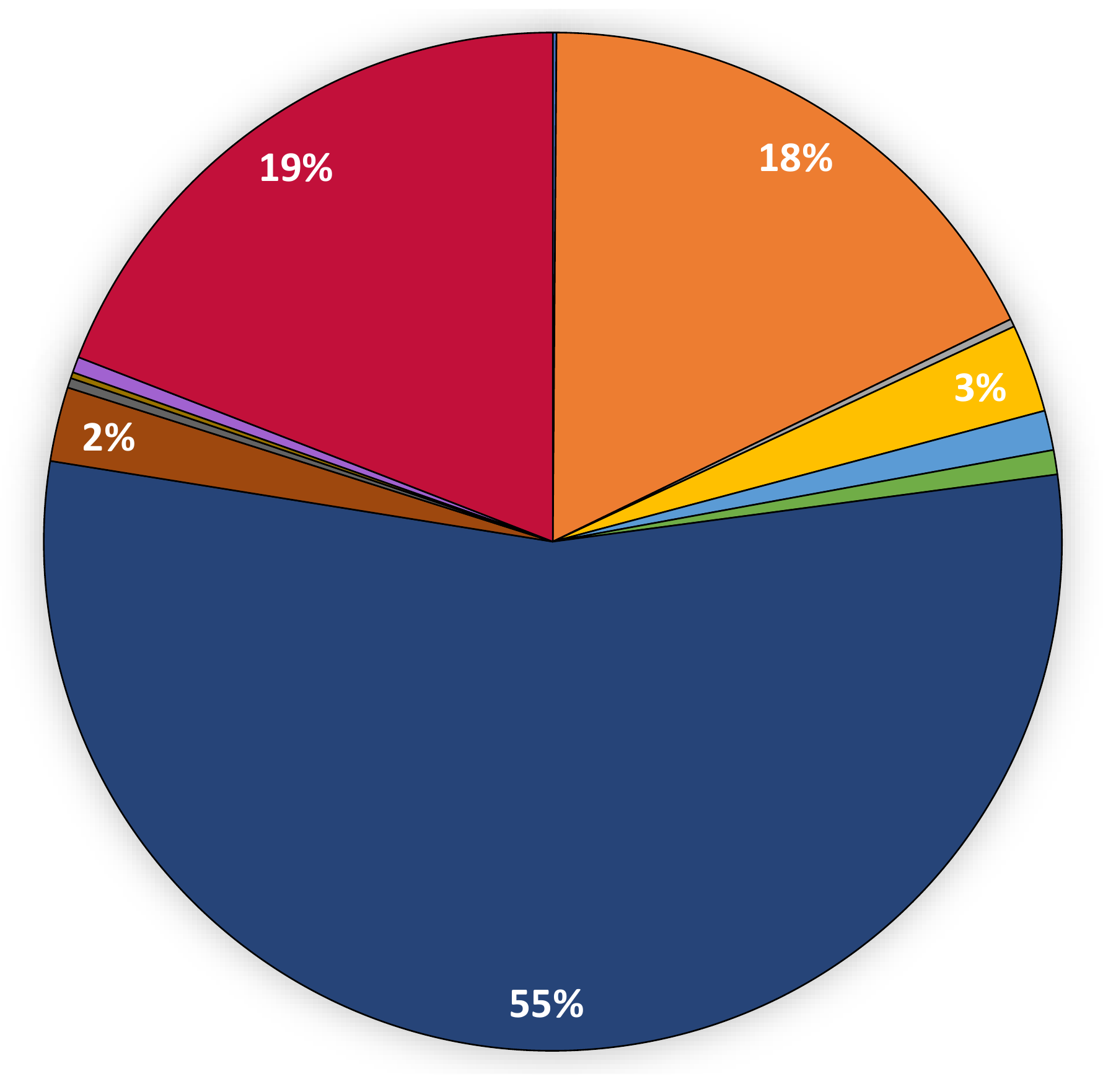}
\caption{Pessimistic}
\label{fig:overall-todo-pes-new}
  \end{subfigure}
\caption{Improved root causes for ACG variants across TodoMVC, after WALA improvements.}
\end{figure*}

\begin{figure}
\centering
\includegraphics[width=\linewidth,height=10cm,keepaspectratio]{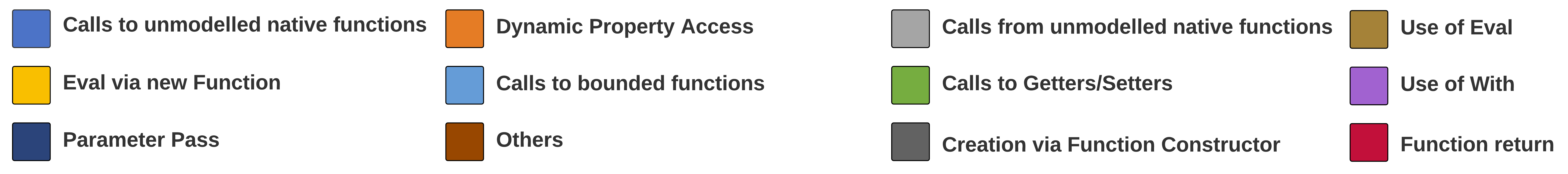}
\end{figure}

\begin{figure}
\centering
\includegraphics[width=\linewidth,height=7cm,keepaspectratio]{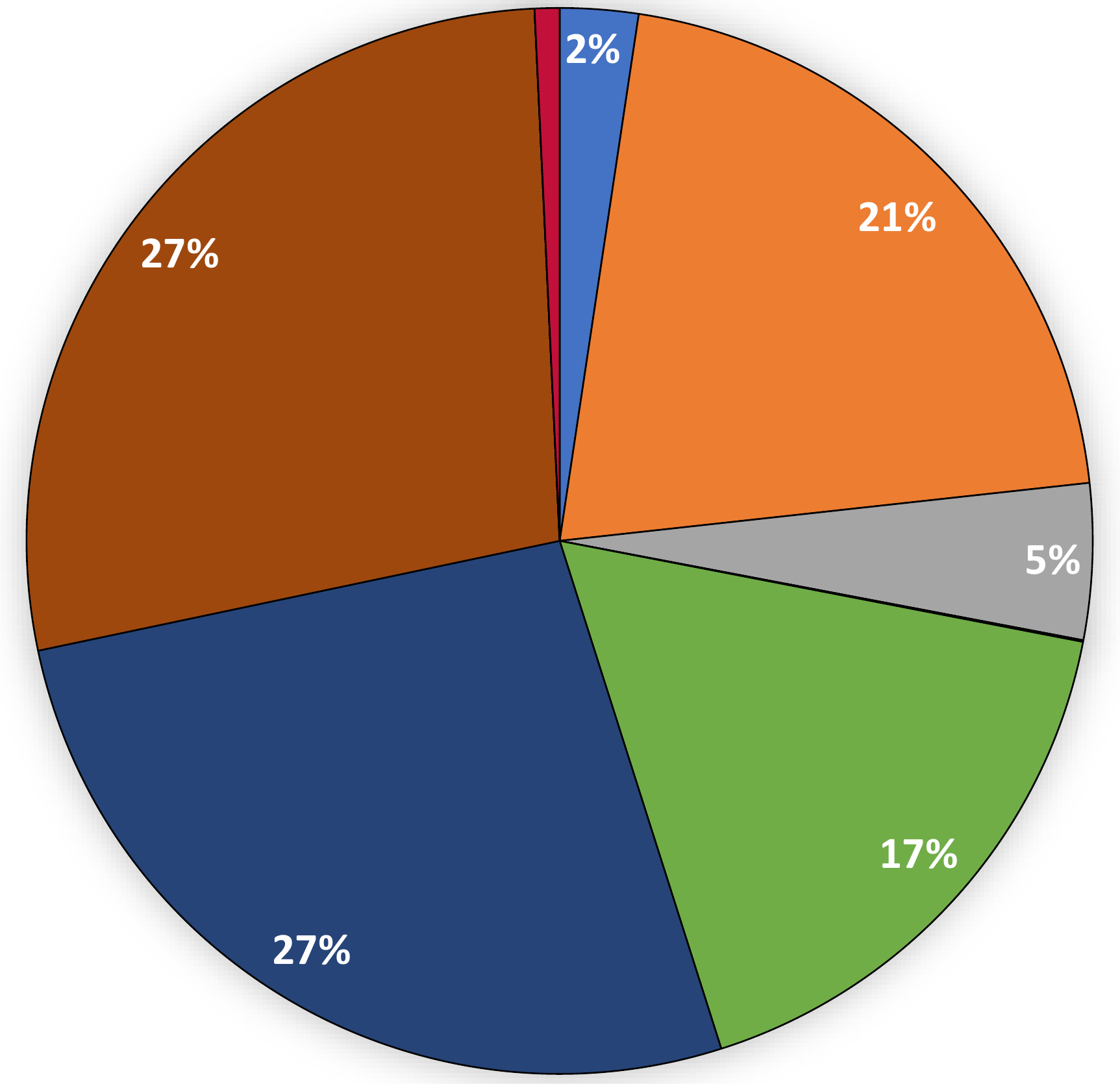}
\caption{Root causes for pessimistic ACG for Juice Shop.}
\label{fig:overall-juiceshop-pes-new}
\end{figure}

\tightpara{Using data to improve recall} \Cref{fig:overall-todo-opt-old} shows the prevalence of different root causes across the TodoMVC benchmarks for the optimistic variant of the original ACG implementation in WALA.  When studying these root causes, we were surprised to see that 24\% of missed call edges were due to calls to unmodeled standard library functions.  Based on this data, we modified WALA to include basic models of many of these native functions. This change improved average recall for the pessimistic analysis by 2 percentage points to 37\% (by the Reachable Edges metric); improvement for optimistic analysis was 5 percentage points, to 76\%.  These improvements show that quantifying root cause prevalence can guide an analysis developer to ``quick wins'' for improving analysis recall.  The data in the remainder of this section were computed using the improved version of WALA ACG.

\tightpara{Top root causes} Turning to RQ3, \Cref{fig:overall-todo-pes-new,fig:overall-todo-opt-new} respectively show top root causes for pessimistic and optimistic ACG across the TodoMVC benchmarks (after improving WALA's native models).  Comparing the two, we see a key difference is that missed calls due to functions being passed as parameters or returned (the ``Parameter Pass'' and ``Function return'' labels) are significant root causes (totaling 74\%) for pessimistic analysis but not optimistic.  This result makes sense, as the key difference between optimistic and pessimistic ACG is that optimistic analysis tracks interprocedural flow of function values.  Given that 74\% of missed edges for pessimistic analysis are due to such interprocedural flows, it seems the best approach to improving pessimistic recall for these benchmarks would be to model some of these flows, rather than attacking other root causes. 

The ``Others'' label covers a small number of cases (5\% overall) where our current scripts cannot yet find a root cause. In addition to the unhandled constructs and cases described in \Cref{sec:implementation}, our automated reasoning failed in rare cases due to a bug in WALA ACG's handling of \code{finally} blocks.  During our work, we identified two other WALA ACG bugs that were fixed by the maintainers. Overall, our techniques successfully handle more than 95\% of the missing call edges for our benchmarks, and we will continue to improve our tools to reduce the number of unhandled cases.

Focusing in on \cref{fig:overall-todo-opt-new}, we see that dynamic property accesses are by far the most prevalent root cause for optimistic analysis of TodoMVC benchmarks at 70\%.  We dig further into these property accesses with a finer-grained labeling in \Cref{sec:property-name-flow}.  The second-most prevalent root cause on average is ``Eval via new Function'' at 10\%, but as we shall see next, the second-highest root cause varies significantly across benchmarks.

\tightpara{Variance across benchmarks} For RQ4, we use illustrative examples to show the variance in root cause prevalence across benchmarks.  \Cref{fig:root-cause-react,fig:root-cause-angular,fig:root-cause-vue} respectively show root causes for the React, Angular, and Vue.js TodoMVC benchmarks, analyzed with optimistic ACG.  While the most-prevalent root cause for each of these benchmarks was dynamic property accesses, the second-place root cause varies by benchmark: ``Eval via new Function'' is second for React, ``Call to bounded functions'' for AngularJS, and ``Call to getter / setter'' for Vue.  This benchmark-specific data could provide valuable information to an analysis developer.  E.g., if the developer were primarily trying to improve recall for applications like the Vue benchmark, it may be more worthwhile to improve handling of getters and setters than if the applications were more similar to the React benchmark.  

\Cref{fig:overall-juiceshop-pes-new} shows root causes for the larger Juice Shop benchmark (analyzed with pessimistic ACG).  Unfortunately, Juice Shop exercised gaps in our infrastructure's handling of tricky JavaScript constructs more heavily, particularly in the dynamic flow trace analysis.  So, we could not compute proper root causes for 27\% of missing call graph edges for Juice Shop.  Still, the remaining data is interesting, particularly when compared to the pessimistic results for the TodoMVC benchmarks shown in~\Cref{fig:overall-todo-pes-new}.  We see that handling returns of functions seems to be relatively less important than for the TodoMVC benchmarks, whereas handling of getters and setters is more important.  Though making strong conclusions is difficult given the number of uncategorized edges in this case, these preliminary data again show the ability of our technique to expose benchmark-specific insights about causes of low recall.

To summarize, we have shown that our technique for quantifying root causes works across several benchmarks and can expose the most important root causes in aggregate and the differences between benchmarks.  Since improving recall for JavaScript static analysis on real-world programs poses so many challenges, we expect improvements for specific types of benchmarks to prove worthwhile, and the data from our techniques can provide valuable guidance in how to do so.

\subsection{Name Flow for Dynamic Property Accesses}\label{sec:property-name-flow}

\begin{figure}
\centering
\includegraphics[width=\linewidth,keepaspectratio]{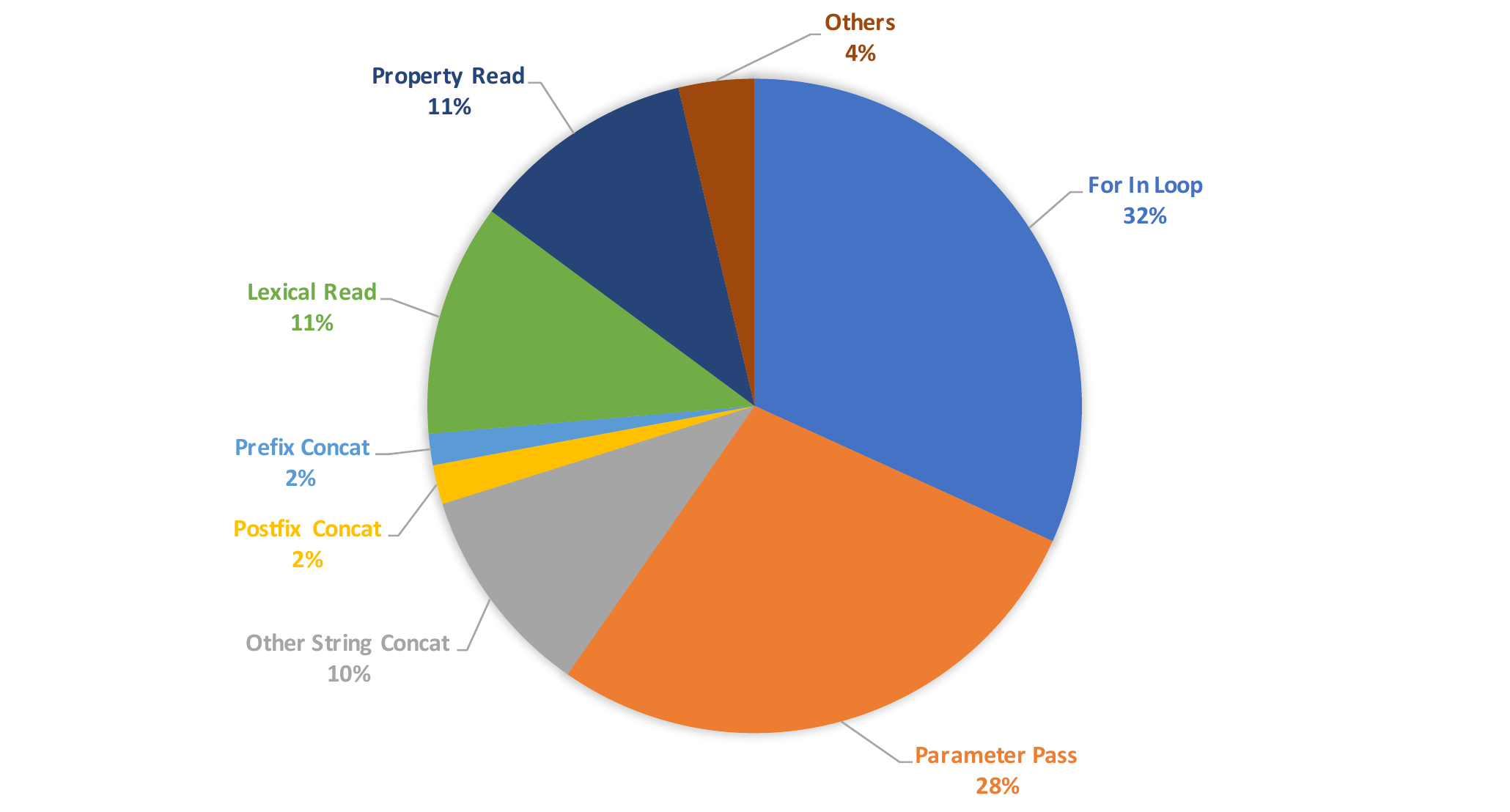}
\caption{Finer-grained dynamic property access root causes for TodoMVC benchmarks.}
\label{fig:property-name-categories}
\end{figure}

Given the importance of dynamic property accesses as a root cause in \Cref{sec:top-root-causes}, we performed a finer-grained root cause labeling of these accesses.  Our goal was to understand better how property names are computed for these accesses, to see if some targeted handling of the property name expressions could be useful.  Recent work by Nielsen et al.~\cite{nielsen21modular} proposes just such a technique for analysis of Node.js code, via special handling of property name expressions that concatenate a string constant prefix or suffix to some other expression.  We hoped to use root cause labeling to see if a similar technique could be effective for our web-based benchmarks.

We implemented a simple intra-procedural analysis using WALA~\cite{wala} to label each root-cause dynamic property access based on how data flows into its property name expression (for an access \code{x[e]}, \code{e} is the property name expression).  Aggregate results appear in \Cref{fig:property-name-categories}; our artifact has the complete data~\cite{acgArtifact}.  As shown in \Cref{fig:property-name-categories}, property names for root-cause dynamic accesses have a diverse set of sources.  The largest single source are JavaScript's \code{for}-\code{in} loops for iterating over object properties, studied frequently in the literature as a challenge for static analysis (e.g.,~\cite{ParkLR18,AndreasenM14}).  However, they account for only 31\% of cases in total, and many other sources exist.  Property names are often passed in from outside the function containing the access, whether by parameter passing (28\%) or variables in enclosing lexical scopes (12\%); handling these cases may require inter-procedural tracking of property name value flow.  Another major source is property reads (12\%) (i.e., the property name is read from another object property), whose handling may again require deep tracking of value flow.

String concatenation cases comprise 14\% of root-cause property name expressions.  Only 4\% of such expressions in our benchmarks had a string constant prefix or suffix, the type of expression targeted by Nielsen et al.~\cite{nielsen21modular}.  Hence, the data show that their technique would likely have at most a small impact on recall for our benchmarks.

A deeper study of inter-procedural property name value flow could provide further insights on how these names are computed; this remains as future work.  Still, our data show it is likely that a variety of challenges would need to be addressed to significantly improve ACG's recall with respect to dynamic property accesses.

\subsection{Threats to Validity}\label{sec:threats-to-validity}

As noted in \Cref{sec:study-setup}, we do not claim generalizability of the results for our benchmarks to a broader set of JavaScript applications.  In our benchmark suite, each individual framework is primarily exercised by a single TodoMVC benchmark, which may not be representative of other applications using that framework.  Also, though our harness achieves high statement coverage for the TodoMVC benchmarks (\Cref{sec:benchmarks-and-harness}), it is possible that certain application behaviors in those apps remain unexercised.  Our dynamic coverage of Juice Shop was relatively low due to scalability limitations; more complete coverage is required to make strong conclusions about relative importance of root causes for that application.  Finally, as noted in \Cref{sec:implementation}, our tooling still does not handle certain language features completely, which may have impacted our measurements.

\section{Related Work}\label{sec:related}

Here, we briefly discuss related studies of analysis effectiveness, and also other analysis frameworks and their applicability to framework-based web applications.

\tightpara{Root cause analysis} Our work was partly inspired by a study of call graph recall for Java programs by Sui et al.~\cite{Sui2020}.  As in that work, we measure recall with respect to dynamic analysis measurements, and we aim to determine which constructs are responsible for missing edges.  Sui et al.'s approach used calling-context trees~\cite{AmmonsBL97} and runtime tagging of reflective operations to determine language features impacting recall.  Since functions are first-class values in JavaScript, we can trace function data flow directly to make this determination.  Also, due to JavaScript's dynamic nature, the potential causes of missing edges and their usage patterns differ significantly from Java's problematic constructs.

Andreasen et al. present techniques for isolating soundness and precision issues in the TAJS static analyzer for JavaScript~\cite{AndreasenMN17}.  For finding analysis unsoundness, their technique creates logs of expression values while executing target programs, and then checks that the static analysis abstractions account for all such values.  When unsoundness is discovered for a program, delta debugging~\cite{DBLP:journals/tse/ZellerH02} is employed to find a reduced version of the program with the same unsoundness.  From this reduced program, determining a root cause is often much simpler.  In contrast to their work, which is focused on an analysis that strives for full soundness, our approach is targeted at analyses with deliberate unsoundness (for practicality), and aims to quantify the impact of different unsoundness root causes.

Reif et al.~\cite{DBLP:journals/tse/ZellerH02} present a system that provides methods for exposing sources of unsoundness in different Java call graph builders and also for measuring how frequently hard-to-analyze constructs appear in a set of benchmarks, yielding many useful practical insights.  A difference with our work is that our technique can automatically connect specific uses of hard-to-analyze constructs to the corresponding missed call graph edges.  This provides important additional information for JavaScript, since hard-to-analyze constructs can appear pervasively in JavaScript code, and not all occurrences cause call graph unsoundness.

Lhot{\'{a}}k~\cite{Lhotak2007} also presents a comparison of static and dynamic
call graphs for Java, aimed at finding sources of imprecision in the static call
graph. Other work~\cite{AndreasenMN17,WeiTRD16} used dynamic analysis to
generate traces and find root causes of imprecision in JavaScript static
analyses, and Wei et al.~\cite{WeiTRD16} also provides suggestions to fix the
root causes of imprecision. Lee et al.~\cite{LeePR20} produce a tracing graph by
tracking information flow from imprecise program points backwards, thereby
aiding the user to identify main causes of the imprecision.  Our work differs
from all of these studies in its focus on recall rather than precision, which
necessitates different techniques.

\tightpara{JavaScript Analyses}
Several analysis frameworks use abstract interpretation~\cite{CousotC77} to
handle the interdependent problem of scalability and precision in
JavaScript~\cite{JensenMT09,Kashyap2014,safe}. These frameworks have been
steadily enhanced with techniques to improve precision and scalability when analyzing
libraries, particularly
TAJS~\cite{JensenMT09,AndreasenM14,Jensen2011,NielsenM19} and
SAFE~\cite{safe,LeePR20,KoRR19,ParkLR13,ParkLR18,RyuPP19}.
While these techniques have shown enormous improvement in analyzing
libraries like jQuery~\cite{jquery} and Lodash~\cite{lodash},
they do not yet scale to complex MVC frameworks
like React~\cite{ReactJS}.

Other techniques use
dynamic information to improve static analysis.  Wei and Ryder introduced
blended analysis~\cite{Wei2012}, which uses dynamic analysis to aid static
analysis in handling JavaScript's dynamic features.  
The dynamic flow analysis by Naus
and Thiemann~\cite{Naus2016} generates flow constraints
from a training run to infer types in JavaScript applications.
(Their technique finds constraints by tracking operations on values; we determine how values are copied through memory, an orthogonal problem.)
Lacuna~\cite{ObbinkMSL18} utilizes static and
dynamic analysis to detect dead code in JavaScript applications; this work uses ACG and also uses TodoMVC applications for evaluation.
While dynamic information can be very helpful in static analysis, 
improving pure static analysis is still desirable, as it can compute results without instrumenting and running the code and without inputs.

To analyze JavaScript applications that use the
Windows runtime and other libraries, Madsen et al. proposed
a use analysis that infers points-to specifications
automatically~\cite{Madsen2013}. It is unclear if their analysis will be effective for framework-based applications, where control flow is mainly driven by the framework, not the application.  Also, we
study applications using diverse frameworks from by many different developers, whereas~\cite{Madsen2013} focuses on Windows libraries. For Node.js, Madsen et al.~\cite{madsen2015static} presented a static analysis using call graphs augmented to represent event-driven control flow.
To scale static analysis in server-side JavaScript applications in Node.js,
Nielsen et al. present a feedback-driven static analysis to automatically
identify the third-party modules that need to be analyzed~\cite{Nielsen2019}.
Our focus, however, is on client-side MVC applications
that often do not have clean module interfaces.

Other recent systems make use of pragmatic JavaScript static analyzers.  The
CodeQL system~\cite{codeqlmain} includes an under-approximate call graph builder for JavaScript~\cite{codeqlcg}.  CodeQL's analysis is primarily intra-procedural, targeted toward taint analysis, and does not handle dynamic property accesses.\footnote{These details are based on personal communication with Max Sch{\"{a}}fer in January 2021.} M{\o}ller et al.~\cite{moller20detecting} describe a system for detecting breaking library changes in Node.js programs, based on an under-approximate analysis designed for high recall at the cost of some precision.  Nielsen et al.~\cite{nielsen21modular} present a pragmatic modular call-graph construction technique for Node.js programs; we discussed its specialized handling of property name expressions in \Cref{sec:property-name-flow}.  For these approaches, our methodology could be used to quantify the importance of different causes of reduced recall.  Salis et al. recently presented a pragmatic call graph builder for Python programs~\cite{salis21pycg}; it would be interesting future work to extend our techniques to Python.  Beyond dataflow-based reasoning about call graphs, other approaches to JavaScript static analysis include AST-based linting~\cite{eslint} and type inference~\cite{ternjs,chandra16typeinf}.

\section{Conclusions}\label{sec:conclusions}

We have presented novel techniques for quantifying the relative importance of different root causes of missed edges in JavaScript static call graphs.  We instantiated our approach to perform a detailed study of the results of the ACG algorithm on modern, framework-based web applications.  The study's results provided numerous insights on the variety and relative impact of root causes for missed edges.  All of our code and data is publicly available.  In future work, we plan to extend the study to other domains; we expect that analyses for any dynamic language with extensive use of higher-order functions could benefit from our techniques.  We also plan to use the techniques to further develop improved call graph builders and other JavaScript static analyses.

\begin{small}
\bibliography{refs}
\end{small}

\iftoggle{extended}{
\clearpage
\appendix

\section{Precision of ACG}\label{sec:precision-details}

In addition to the recall measurements of \Cref{sec:recall-measurements}, we also measured precision of the pessimistic and optimistic variants of ACG on our benchmarks.  \Cref{fig:avg-recall-precsion-b4} gives aggregate precision and recall data for the TodoMVC benchmarks (we include the recall data again for easy side-by-side comparison), while \Cref{fig:pes-precsion-b4,fig:opt-precsion-b4} respectively give per-benchmark precision data for optimistic and pessimistic variants.  We observe significant differences in results for optimistic vs. pessimistic analysis, depending on the metric used.

 Consistent with Feldthaus et al.~\cite{Feldthaus2013}, our data in \Cref{fig:avg-recall-precsion-b4} show that with the call site targets metric, precision and recall of optimistic and pessimistic ACG tend to be similar (recall for React is an exception, as shown previously in \Cref{fig:recall-results-detailed}).  With reachable nodes, optimistic ACG tends to have significantly higher average recall (80.9\% vs. 43.2\%) with only slightly lower average precision (51.9\% vs 53.0\%).  Hence, for reachable nodes, optimistic ACG provides better results than pessimistic ACG for our benchmarks.  For the reachable edges metric, optimistic ACG provides a large increase in average recall (69.6\% vs 29.4\%), but with a large reduction in average precision (11.7\% vs. 22.4\%).  Hence, on these benchmarks, neither ACG variant provides a very accurate call graph in terms of the reachable edges metric.  In summary, the relative merits of optimistic vs. pessimistic ACG in terms of precision and recall are highly dependent on what metrics are used for measurement.
 
\begin{figure}[t]
  \centering
  \includegraphics[width=0.75\linewidth,keepaspectratio]{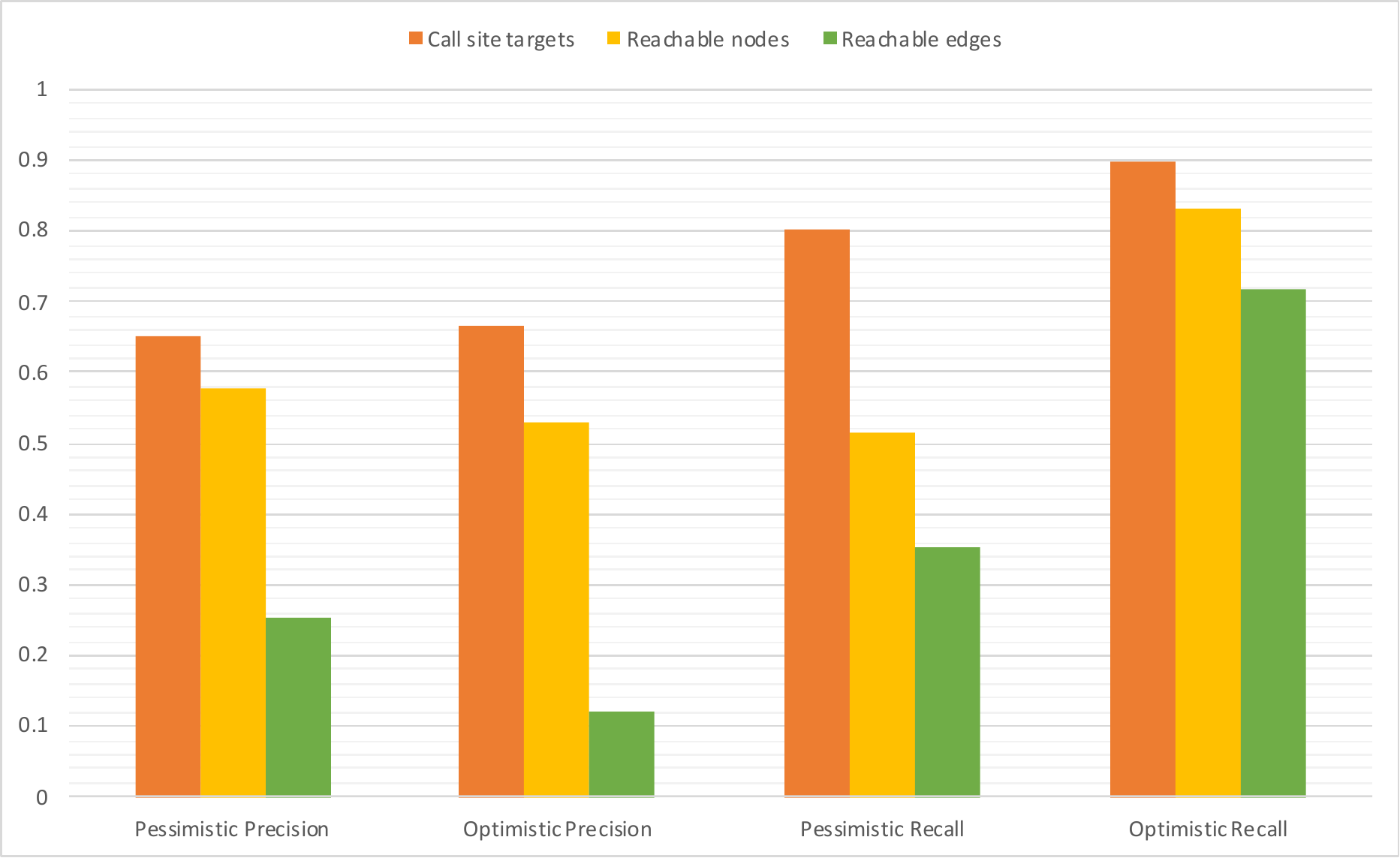}
  \caption{Average precision and recall across TodoMVC benchmarks.}
  \label{fig:avg-recall-precsion-b4}
\end{figure}

\begin{figure}[t]
  \centering
  \includegraphics[width=0.75\linewidth,keepaspectratio]{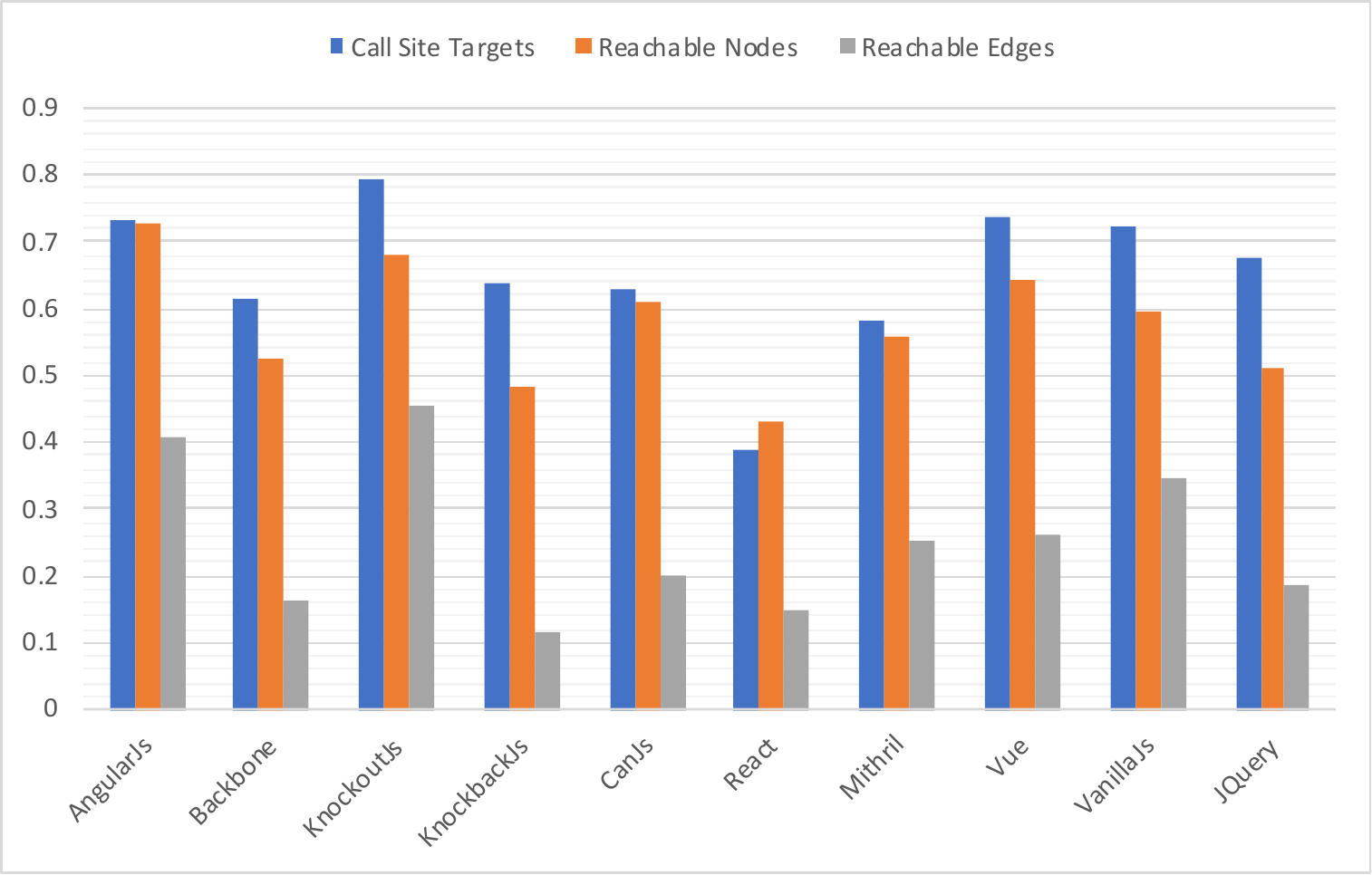}
  \caption{Pessimistic precision across TodoMVC benchmarks.}
  \label{fig:pes-precsion-b4}
\end{figure}

\begin{figure}[t]
  \centering
  \includegraphics[width=0.75\linewidth,keepaspectratio]{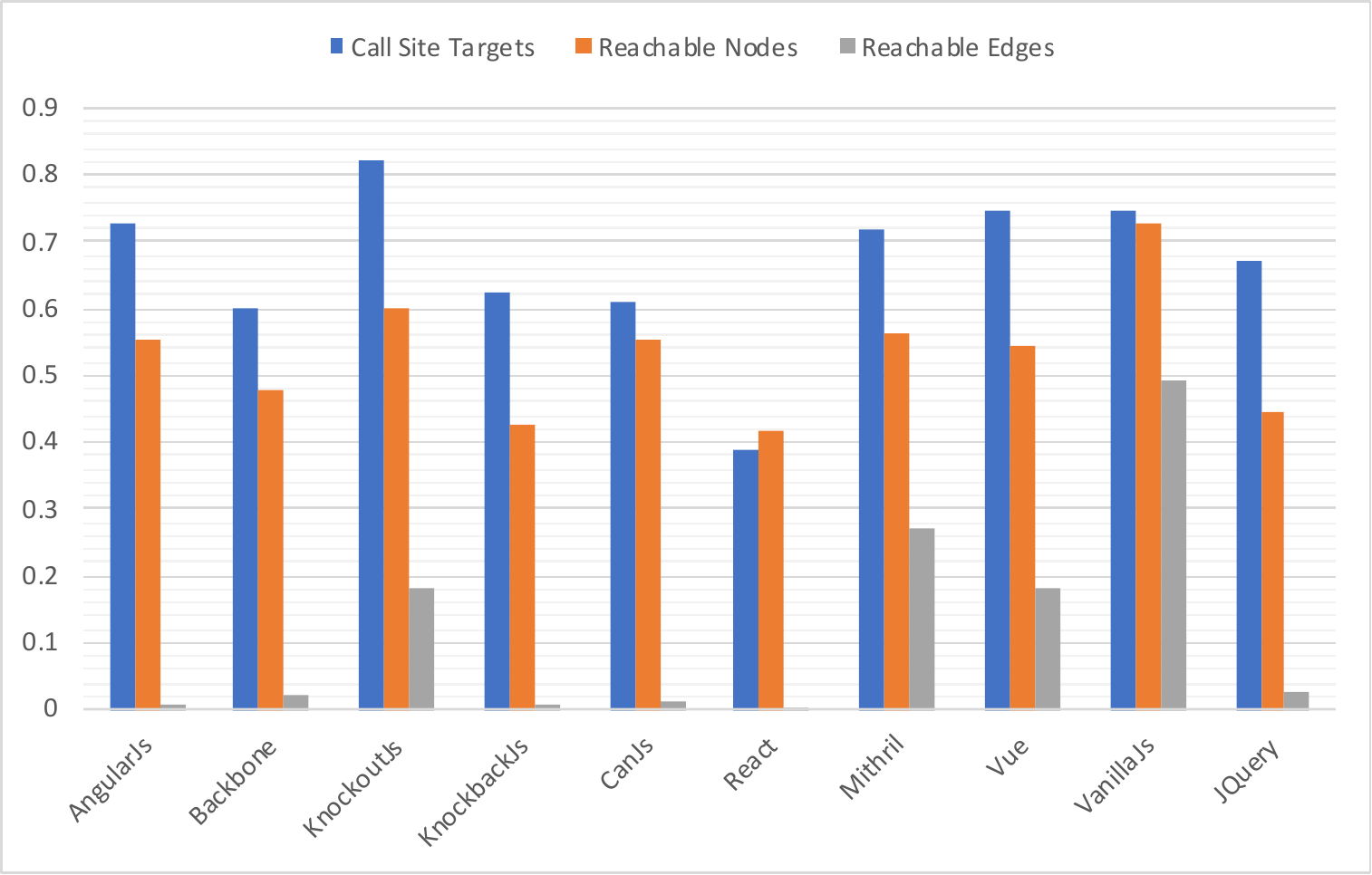}
  \caption{Optimistic precision across TodoMVC benchmarks.}
  \label{fig:opt-precsion-b4}
\end{figure}
}{}
\end{document}